\documentclass[amssymb,amsmath,floatfix,aps,pre,twocolumn,superscriptaddress,longbibliography,nofootinbib]{revtex4-2}
\usepackage[latin1]{inputenc}

\usepackage{mathptmx}
\usepackage[T1]{fontenc}

\usepackage{graphicx}
\usepackage{amsmath}
\usepackage{amssymb}
\usepackage{amsfonts}
\usepackage[scr=rsfs]{mathalpha}
\usepackage{bm}
\usepackage{hyperref}
\usepackage{bbold}
\usepackage{xcolor}
\usepackage{overpic}
\usepackage{siunitx}

\newcommand{\algn}[1]{\begin{align} #1 \end{align}}
\newcommand{\sbeqs}[1]{\begin{subequations} #1 \end{subequations}}
\newcommand{\pmat}[1]{\begin{pmatrix} #1 \end{pmatrix}}
\newcommand{\css}[1]{\begin{cases} #1 \end{cases}}

\newcommand{\eps}{\ensuremath{\varepsilon}}
\newcommand{\ve}[1]{\boldsymbol{#1}}
\newcommand{\nn}{\nonumber}
\newcommand{\ee}{\ensuremath{\text{e}}}
\newcommand{\ed}{\ensuremath{\text{d}}}
\newcommand{\dd}[1]{\ensuremath{\frac{\text{d}}{\text{d} #1}}}
\newcommand{\ddd}[1]{\ensuremath{\frac{\text{d}^2}{\text{d} #1^2}}}
\newcommand{\deldel}[1]{\ensuremath{\frac{\partial}{\partial #1}}}

\newcommand{\mc}[1]{\ensuremath{\mathcal{#1}}}
\newcommand{\ms}[1]{\ensuremath{\mathscr{#1}}}
\newcommand{\mbb}[1]{\ensuremath{\mathbb{#1}}}
\newcommand{\kb}{\ensuremath{k_\text{B}}}
\newcommand{\eqnlab}[1]{\label{eq:#1}}
\newcommand{\seclab}[1]{\label{sec:#1}}
\newcommand{\figlab}[1]{\label{fig:#1}}
\newcommand{\eqnref}[1]{\eqref{eq:#1}}
\newcommand{\Eqnref}[1]{Eq.~\eqref{eq:#1}}
\newcommand{\Eqsref}[1]{Eqs.~\eqref{eq:#1}}
\newcommand{\secref}[1]{\ref{sec:#1}}
\newcommand{\Secref}[1]{Sec.~\ref{sec:#1}}
\newcommand{\figref}[1]{\ref{fig:#1}}
\newcommand{\Figref}[1]{Fig.~\ref{fig:#1}}
\newcommand{\Figsref}[1]{Figs.~\ref{fig:#1}}

\DeclareMathOperator{\arccot}{arccot}

\DeclareMathOperator{\arccoth}{arccoth} 

\newcommand{\Jmat}{\ensuremath{\mathbb{J}}}       % symplectic matrix
\newcommand{\Hmat}{\ensuremath{\mathbb{H}}}       % Hamiltonian matrix
\newcommand{\Cmat}{\ensuremath{\mathbb{C}}}       % cost matrix
\newcommand{\Bmat}[1]{\ensuremath{\mathbb{B}^{#1}}} % boundary matrices
\newcommand{\Smat}{\ensuremath{\mathbb{S}}}       % canonical transformation matrix
\newcommand{\wvec}{\ensuremath{\boldsymbol{w}}}       % canonical phase-space vector (u, mu)
\newcommand{\vvec}{\ensuremath{\boldsymbol{v}}}       % state-control vector (u, lambda)

\begin{document}

\title{Equivalence classes of finite-time transitions in optimal control and non-equilibrium relaxation}

\author{Jan Meibohm}
\affiliation{Institute for Physics and Astronomy, Technische Universit\"at Berlin,
  Hardenbergstra\ss{}e 36, 10623 Berlin}
\author{Samuel Monter}
\affiliation{Faculty of Physics, University of Konstanz, Konstanz, Germany}
\author{Clemens Bechinger}
\affiliation{Faculty of Physics, University of Konstanz, Konstanz, Germany}
\author{Sarah A. M. Loos}
\affiliation{Max Planck Institute for Dynamics and Self-Organization, G\"ottingen}

\begin{abstract}
We present a theory for the optimal control of stochastic systems in structured environments,
represented by penalty terms in the cost functional. We show that such control problems generically feature sharp finite-time transitions associated with a qualitative change in the control strategy at a critical time. Starting from an overdamped Langevin equation
and a quadratic cost functional, we show that all resulting problems fall into three canonical
equivalence classes (parabolic, hyperbolic, and elliptic), distinguished by the sign of the determinant of the
control Hamiltonian. For each class, we
obtain the optimal protocol, the cost function, and the critical time in closed
form, and show that the transition exhibits features of a continuous phase transition at mean-field level. We then establish a mapping between the optimal control cost and the large-deviation rate function governing non-equilibrium relaxation
after a potential quench. The mapping covers the parabolic and hyperbolic classes,
while the elliptic class has no simple relaxation counterpart. This correspondence implies that recently discovered
finite-time dynamical phase transitions, which are exponentially costly to
sample directly, are accessible through ordinary averages over optimally controlled
trajectories. To validate our theoretical findings, we report three experiments with optically trapped colloidal
particles: a control transition for the mean stochastic work, and the finite-time
dynamical phase transitions in free diffusion and in harmonic relaxation.
\end{abstract}

\maketitle

% =========================================================
\section{Introduction}
\seclab{intro}
% =========================================================

With the ongoing miniaturization of technological devices and the emergence of synthetic
microrobots, the optimal control of stochastic processes has become a central problem at the
intersection of statistical physics, soft matter, and biophysics~\cite{Chi24,Ju25}. At the
mesoscale, thermal fluctuations are comparable in magnitude to the
relevant systematic forces at play,
and frictional forces cause large dissipative losses. Biological systems, e.g., molecular motors, flagellated bacteria, and eukaryotic
cells, have evolved to operate efficiently under these conditions, completing tasks using finite resources and within
finite time windows despite strong noise~\cite{How01,Jul97,Sog17}. A fundamental feature of
all finite-time processes is the trade-off between speed and energetic cost: faster protocols
incur higher irreversible dissipation, while slower protocols reduce energetic cost at the
expense of throughput. Optimal control theory, which identifies protocols minimizing a
prescribed cost functional over a finite time horizon, provides the natural quantitative
framework to optimize this trade-off under given constraints, and has become an indispensable tool for understanding and
designing mesoscopic machines~\cite{Fle75,Bec21,Alv25}.

Already in simple situations, optimal finite-time protocols can become highly non-trivial. For example, for an
overdamped colloidal particle dragged by a harmonic trap, the minimum work protocol features
finite discontinuities at the temporal boundaries~\cite{Sch07}. 
Recent works have explored how intrinsic activity~\cite{Gup23}, dynamic memory~\cite{Loo24}, or the presence of particle interactions~\cite{Mon26} add further non-trivial nonlinearities~\cite{Ols25} and discontinuities~\cite{Wel26} to the optimal solutions.
Here, we turn our attention to structured environments, where the particle's position 
is associated with a spatially varying (e.g. energetic) cost, which is a common situation in many real-world scenarios, including a particle navigating past an obstacle [\Figref{overview}(a)], a microswimmer steering through a tight constriction~\cite{Loo24,Col17,Gar25,Gup23}, or a molecular motor avoiding a
high-energy conformation~\cite{How01,Jul97}. Such energy landscapes turn out to dramatically affect the control problem. We analyze the effect of structured environments for the minimum case, where only the particle's endpoint is subject to a space-dependent penalty. Consequently, the final position becomes itself part of the optimization, and its optimal value reflects a competition between path-dependent cost and position-dependent penalty. As we demonstrate in a companion letter~\cite{Mei26b}, for a particle to be transported in a harmonic trap at minimum mean work, this
competition produces a sharp finite-time transition in the optimal control strategy at a
critical duration $t_\text{c}$: for short protocols the particle accepts the positional penalty,
whereas for long protocols it actively steers toward regions of lower cost. When the penalty
landscape is symmetric, this transition is accompanied by spontaneous symmetry breaking of
the optimal final position~\cite{Kap05a,Kap05b}, in direct analogy with a continuous phase
transition. First-order transitions have been observed in optimal control in other contexts~\cite{Sol18}.

In this paper, we develop the theoretical framework underlying the results
of~\cite{Mei26b} and extend them substantially. Starting from a nonlinear overdamped
Langevin equation, we derive a universal linearized control problem by expanding the dynamics
about the ``zero-control fixed point''. This point is given by the stationary solution of the noise-free dynamics
under the zero-control protocol, defined precisely in \Secref{nonlinear}. We show that, up
to a class of canonical transformations, all such linearized problems fall into one of three
equivalence classes [\Figref{overview}(b)] distinguished by the sign of the determinant $\det\Hmat$ of the control
Hamiltonian matrix $\Hmat$: the parabolic class ($\det\Hmat = 0$) treated in~\cite{Mei26b}, the hyperbolic class
($\det\Hmat < 0$), and the elliptic class ($\det\Hmat > 0$). These are the same three classes
that classify linear Hamiltonian systems in classical mechanics~\cite{Wil36,Arn01}, and they carry
the same mechanical picture: free motion, motion in an inverted parabola, and harmonic
oscillation, respectively. However,
the physical meaning of $\det\Hmat$ is specific to the control problem analyzed here: $\det\Hmat$ characterizes the response of the control cost to mutual shifts of the mean coordinate and the control. For each class, we derive explicit analytical expressions for the optimal protocols, cost functions, and critical times at which control transitions occur. The ``harmonic-trap, mean-work problem'' of~\cite{Mei26b} is
the canonical representative of the parabolic class. In particular, we show that all parabolic problems are characterized
by homogeneity of space, i.e., they are dynamically invariant under the combined shift of the mean coordinate and the control,
and can be reduced to the mean-work problem of Ref.~\cite{Mei26b} by a canonical transformation.

\begin{figure*}[t]
    \centering
    \includegraphics[width=.8\linewidth]{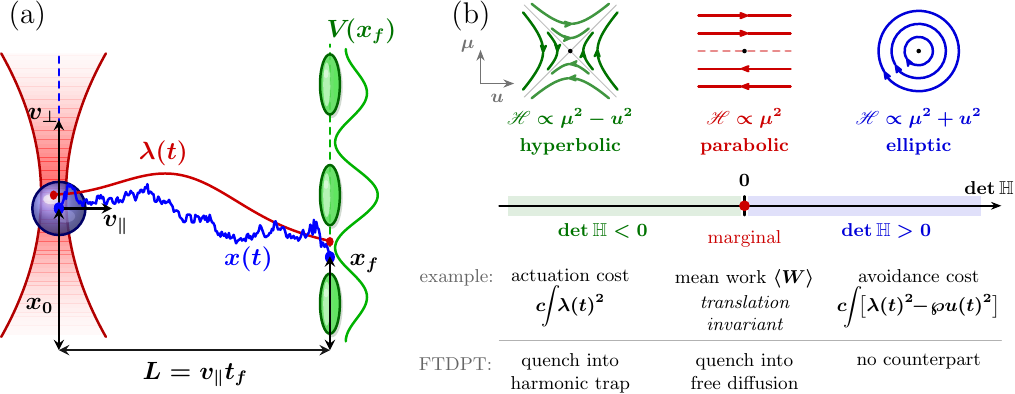}
    \caption{(a) Control of a Brownian particle (blue), initially at $x_0$ and moving with constant mean velocity $v_\parallel$ towards an obstacle (green) at distance $L=v_\parallel t_f$. An optical trap (red) with center $\lambda(t)$ guides the particle's vertical position $x(t)$ (blue) via $v_\perp$ over the duration $t_f$. At the end, the particle must cross the obstacle described by the penalty $V(x_f)$, at the final position $x_f=x(t_f)$. (b) The three
    equivalence classes, distinguished by the sign of $\det \mbb{H}$, i.e., the determinant of the control Hamiltonian \eqnref{Hquad} [see \Secref{canonical}]. Top:
    Hamiltonian flow in the $(u,\mu)$ plane, with the normal form \eqnref{H0normal} of
    the control Hamiltonian $\ms{H}$. Bottom: Representative cost functionals for each class, and
    corresponding relaxation process whose FTDPT it maps onto. The mean
    work is marginal, $\det \mbb{H}=0$, as a consequence of the translation
    invariance of the work functional and the dynamics.}
    \figlab{overview}
\end{figure*}

Another central theme of this paper is the connection between the optimal control
transition and \textit{finite-time dynamical phase transitions (FTDPTs)} in non-equilibrium
relaxation after a potential quench~\cite{Mei22a,Blo22,Mei23b,Vad24,Asr26}, see also~\cite{Ent02,Kul07,Erm10,Fer13}. This connection is rooted in the equivalence between the Pontryagin minimum principle and the saddle-point approximation of path
integrals~\cite{Fle06,Tou09,Che15b,Gra19}. Through this link, the optimal control cost maps onto the large-deviation
rate function governing rare relaxation trajectories of a stochastic system. The mapping
involves a scaled time reversal, which exchanges initial and final boundary conditions, and a harmonic shift of the terminal
cost. The finite-time
control transition then corresponds to an FTDPT, i.e., a sudden change in the dominant
relaxation pathway at a critical time. This mapping holds for the parabolic class (free
relaxation) and the hyperbolic class (relaxation into a harmonic potential). For the
elliptic class, we find that there is no direct relaxation counterpart.
A key practical implication is that for the parabolic and hyperbolic classes, optimal control experiments access the same
rare-event information as the FTDPT through statistical averages over controlled
trajectories, circumventing the exponential sampling cost that makes direct observation of an FTDPT
challenging~\cite{Buc04}.

We complement the theory with three experiments on optically trapped colloidal
particles moving past a double-well obstacle. The first realizes the control
transition of the parabolic class directly, by implementing the optimal
protocol and measuring the mean work. The second and third measure the
associated FTDPTs in relaxation, i.e., free diffusion for the parabolic class, and relaxation into a harmonic trap for the
hyperbolic class. The first two are reported in the companion
Letter~\cite{Mei26b} and summarized here, while the harmonic relaxation experiment is
presented in detail here. For the measurements of large deviations, we reconstruct the associated rate
function by reweighting an ensemble of unconditioned relaxation trajectories, rather than sampling the rare events directly.
This brings the measurement into reach and resolves the growth and scaling
collapse of the order-parameter susceptibility.

The remainder of this paper is organized as follows. \Secref{model} introduces the general
model, derives the linearized Langevin equation and cost functional, and establishes the
notation used throughout. \Secref{hamiltonian} develops the Hamiltonian formulation of the
optimal control problem and derives the three canonical classes via canonical transformations.
\Secref{transition} analyzes the finite-time control transition, the Landau analogy, and the
optimal protocols for each class. \Secref{mapping} establishes the mapping to non-equilibrium
relaxation, identifies the key differences between the two problems, and explains the elliptic class and its peculiarity. \Secref{experiments} presents the three experiments that we conducted. We discuss
implications and open problems in \Secref{discussion} and summarize in \Secref{conclusion}.
Additional details on derivations and experimental methods are collected in the appendices.

% =========================================================
\section{Model}
\seclab{model}
% =========================================================
We aim to describe the optimal control of a noisy system in an inhomogeneous environment.
The dynamics of the system is described by an overdamped Langevin equation, where the external control
parameter enters explicitly. Actively controlling the particle has an associated cost, described by
a cost function. The inhomogeneous environment enters through an additional endpoint cost,
that penalizes the final value of the coordinate. We now discuss these ingredients in detail.
\subsection{Equation of motion}
\seclab{nonlinear}
We consider a particle whose one-dimensional coordinate $x(t)$ evolves in the time window
$0\leq t\leq t_f$ under the overdamped Langevin equation
\algn{\eqnlab{Langevin}
    \gamma\,\dot x(t) = h[x(t),\lambda(t)] + \sqrt{2g[x(t),\lambda(t)]}\,\xi(t)\,.
}
Here, $\gamma$ is a friction coefficient (assumed to be constant in time and space), $\lambda(t)$ is the control parameter, $h(x,\lambda)$ is the deterministic drift, $g(x,\lambda)\geq 0$ is the (possibly state-dependent) noise strength, and $\xi(t)$ is Gaussian white noise
satisfying $\langle\xi(t)\rangle = 0$ and $\langle\xi(t)\xi(t')\rangle = \delta(t-t')$, stemming from the coupling to an equilibrium environment.

The class of dynamics described by \Eqnref{Langevin} covers a range of physical systems. The generalization higher dimensional $x(t)$ and $\lambda(t)$ is briefly discussed in \Secref{discussion}.

\paragraph*{Zero-control fixed point.} Equation~\eqnref{Langevin} is in general non-linear with multiplicative noise and thus difficult solve. To simplify it, we assume that there exists a \emph{zero-control protocol}, i.e., a constant protocol $\lambda(t) = \lambda^*$ for all
$t\in[0,t_f]$, under which the noise-free dynamics $\gamma \dot x = h(x,\lambda^*)$ has a stable fixed
point at $x = x^*$, i.e.,
\algn{\eqnlab{fixedpoint}
    h(x^*,\lambda^*) = 0\,, \quad \partial_x h(x^*,\lambda^*)<0\,.
}
The zero-control protocol represents the natural operating point of the system in the absence
of active steering. Without loss of generality, we shift the position and control coordinates
to place the fixed point at the origin,
\algn{
    x \to x - x^*\,,\qquad \lambda \to \lambda - \lambda^*\,,
}
so that in the shifted coordinates, $ h(0,0) = 0$ and the zero-control protocol is
$\lambda(t) = 0$ for all $t$. All quantities below refer to these shifted coordinates.

\paragraph*{Linearization.} We expand $h(x,\lambda)$ about the fixed point to first order,
\algn{\eqnlab{driftexpand}
    h(x,\lambda) \sim h_{10}\,x + h_{01}\,\lambda\,,
}
where the linearization coefficients are
\algn{\eqnlab{lincoeffs}
    h_{10} \equiv \partial_{x} h(0,0)<0\,,\qquad
    h_{01} \equiv \partial_{\lambda} h(0,0)\,.
}
The coefficient $h_{10}$ characterizes the stability of the fixed point under the
zero-control dynamics, which is assumed to be stable ($h_{10}<0$). The coefficient $h_{01}$ quantifies the responsiveness of the drift to variations in the control, and can have either sign.

We similarly approximate the noise amplitude at its value at the fixed point,
\algn{\eqnlab{Dconst}
    \sqrt{2g(x,\lambda)} \sim \sqrt{2\Gamma}\,,\qquad \Gamma \equiv g(0,0)\,,
}
treating the diffusion coefficient as a constant throughout the protocol. For $\Gamma>0$, this approximation
is internally consistent with the linearization \eqnref{driftexpand}. Systems for which $\Gamma = 0$, i.e., where the noise amplitude vanishes at the fixed
point, require separate treatment, since multiplicative corrections can then enter at leading
order. We exclude such cases from the present analysis.

After the linearization and constant-diffusion approximation, the equation of motion becomes
\algn{\eqnlab{LEQlin}
    \gamma\dot x(t) = -\partial_x U[x(t),\lambda(t)] + \sqrt{2\kb T/ \gamma}\,\xi(t)\,,
}
for $t\in[0,t_f]$ with the local harmonic potential
\algn{
	U(x,\lambda) = \frac{\kappa}{2}(x-\zeta \lambda)^2,
}
where $\kappa = |h_{10}|$ and $\zeta = h_{01}/|h_{10}|$, and where the fluctuation-dissipation theorem requires that $\Gamma=\kb T\gamma$. Furthermore, since the control $\lambda$ is an external quantity, we can absorb $\zeta$ into $\lambda$ by imposing $\lambda\to\zeta^{-1}\lambda$. This way, the dynamics~\eqnref{LEQlin} reduces to the dynamics of a Brownian particle trapped by a harmonic potential
$U(x,\lambda) = \frac{\kappa}{2}(x-\lambda)^2$, where $\kappa$ is the trap stiffness and
$\lambda(t)$ the time-dependent trap center, see \Figref{overview}(a) for an illustration. The linearized Langevin equation reads
\algn{\eqnlab{LangevinHarm}
    \dot x(t) = -\tau_\text{p}^{-1}[x(t) - \lambda(t)] + \sqrt{2\kb T/\gamma}\,\xi(t)\,,
}
with relaxation time $\tau_\text{p} = \gamma/\kappa$ and friction coefficient $\gamma$. This
equation of motion readily yields an accurate model of our primary experimental system, a colloidal particle trapped by optical tweezers~\cite{Mei26b}. The drift in \Eqnref{LangevinHarm} is already linear in $x$ and $\lambda$, so the linearization \Eqnref{driftexpand} is exact. The fixed
point at the origin corresponds to the particle resting at $x^*=0$ with the trap centered at
$\lambda^*=0$. Hence, after the aforementioned redefinition of $\lambda$, \Eqnref{LangevinHarm} is identical to the linearized Langevin equation \Eqnref{LEQlin} close to a stable fixed point.

 Equation~\eqnref{LangevinHarm} is a linear Langevin equation with additive noise. Averaging over the thermal noise,
\Eqnref{LEQlin} decouples the dynamics of the mean trajectory $u(t)\equiv\langle x(t)\rangle$
from the variance $\sigma^2(t)\equiv\langle x^2(t)\rangle - u^2(t)$. The mean then satisfies
\algn{\eqnlab{meandyn}
   \dot u(t) = -\tau_\text{p}^{-1}[u(t) - \lambda(t)]\,,
}
and the variance obeys $\gamma\dd{t}\sigma^2 = -2\kappa\sigma^2 + 2\kb T$, independently of the
protocol. Hence, the variance relaxes to the stationary value
$\sigma^2_\text{st} = \kb T/\kappa$ as time evolves.
%
% -------------------------------------------------------
\subsection{Quadratic cost functional}
\seclab{cost}
% -------------------------------------------------------
After having discussed the dynamics of the particle position $x(t)$, we now describe the control of the system and its associated cost. The aim of the control is to minimize the mean total cost over both the
protocol $\lambda(t)$, $t\in[0,t_f]$, and the final mean position
$u_{f} \equiv u(t_f)$, for fixed initial condition $u(0) = u_0$.
The optimization over $u_{f}$ reflects the competition between the cost of the control protocol necessary to reach the final point and the endpoint cost itself, determined by the obstacle.

We write the mean total cost in the general quadratic form
\algn{\eqnlab{cost}
    \ms{C}_{t_f}[u(t),\lambda(t)] = \int_0^{t_f}\!\!\ed t\,\vvec^{\sf T}(t)\,\Cmat\,\vvec(t)
    + F(\vvec_f;\vvec_0) + \langle V(x(t_f))\rangle\,,
}
where $\vvec(t) = [u(t),\lambda(t)]^{\sf T}$ is the state-control vector,
$\vvec_{f,0} = [u_{f,0},\lambda_{f,0}]^{\sf T}$ denotes its values at the
endpoints, $\Cmat$ is a symmetric cost matrix with $C_{22}>0$,
and the boundary function
\algn{\eqnlab{boundary}
    F(\vvec_f;\vvec_0) = \vvec_f^{\sf T}\Bmat{f}\vvec_f - \vvec_0^{\sf T}\Bmat{0}\vvec_0
}
contains symmetric boundary matrices $\Bmat{f}$ and $\Bmat{0}$.
The final term $\langle V(x(t_f))\rangle$ is the
noise-averaged penalty at the final position $x(t_f)$. It is kept general for now, and is used here to model inhomogeneities of the environment, e.g., an obstacle at a certain distance. Such a situation is shown in~\Figref{overview}(a), where a particle drifts towards an obstacle at distance $L$ with a fixed mean velocity $v_\parallel$, while the particle velocity $v_\perp$ in the perpendicular direction is controlled. The final position $x_f$ at which the particle collides with the obstacle, modeled by $V$, determines the endpoint cost $V(x_f)$. In the general context of control theory, such endpoint terms arise from penalizing how far the system deviates from the desired target state.

\paragraph*{Noise-averaged penalty.}
Because $x(t)$ is Gaussian under the linearized dynamics \Eqnref{LEQlin},
its distribution at time $t$ is fully characterized by its mean $u_f$ and
variance $\sigma^2(t)$. We assume that $\sigma^2(t)\approx\sigma^2_\text{st}=\kb T/\kappa$, which is justified when the process starts at, or close to, equilibrium and the protocol duration $t_f$ is not too short compared to $\tau_\text{p}$.
In this case, the noise-averaged penalty $\langle V(x(t))\rangle$,
\algn{\eqnlab{Vtildedef}
   \langle V(x(t))\rangle =  \tilde V(u_f)
    = \int_{-\infty}^{\infty}\!\!\!\ed x\,
    \frac{\ee^{-(x-u_f)^2/(2\sigma^2_\text{st})}}{\sqrt{2\pi\sigma^2_\text{st}}}\,V(x)
}
is a function of $u_f$ alone. However, the effective penalty $\tilde V$ generally
differs from $V$ due to the thermal smearing by $\sigma^2_\text{st}$.
For the specific double-well penalty used in the experiments~\cite{Mei26b},
\algn{\eqnlab{dwell-main}
	V_\text{DW}(x) = \frac{V_0}4\left[\left(\frac{x}{x_\text{m}}\right)^2-1\right]^2
}
 the noise-averaged form reads explicitly (derived in Appendix~\secref{Apotential})
\algn{\eqnlab{VtildeDW}
    \tilde V_\text{DW}(u_f) = V_\text{DW}(u_f) + \frac{\eps V_0}{4}
    \left(\frac{6u_f^2}{x_\text{m}^2} - 2 + 3\eps\right)\,,
}
where we introduced the dimensionless parameter 
\algn{\eqnlab{epsdef}
	\eps = \frac{\kb T}{\kappa x_\text{m}^2}\,,
}
which is a measure of the level of thermal noise. Figure~\figref{potential} shows the effective obstacle potential $\tilde V_\text{DW}$ as a function of $u$ for different noise levels $\eps$. For weak noise, $\eps\ll1$, i.e., $\kb T\ll \kappa x_\text{m}^2$ and $\sigma^2_\text{st}\ll x_\text{m}^2$, we have $\tilde V(u_f)\sim V(u_f)$. However, stronger noise may qualitatively change the noise-averaged potential $\tilde V$. In particular, for strong noise ($\eps>1/3$), the coefficient of $u_f^2$ in $\tilde V_\text{DW}$ changes sign, so the double well is smeared into a single well. In this case, due to the high noise level, the smallest penalty is generated for $u_f=0$, because moving closer to the minima $\pm x_\text{m}$ of the bare $V_\text{DW}$ increases the probability of being driven into the large wings of $V_\text{DW}$, thus increasing the effective $\tilde V(u_f)$ for $u_f\neq0$.

Overall, we obtain a one-dimensional optimal control problem, with a general quadratic cost~\eqnref{cost} in $u(t)$ and $\lambda(t)$  and with a terminal cost $\tilde V(u_f)$. The quadratic cost is consistent with the linearized dynamics~\eqnref{meandyn} and follows from a Taylor expansion of a smooth cost functional in the displacement from the zero-control fixed point, retaining
terms up to second order. The constraint $C_{22}>0$ ensures that the running cost penalizes large control amplitudes and that the problem is well-posed.
\begin{figure}[t]
    \centering
    \includegraphics{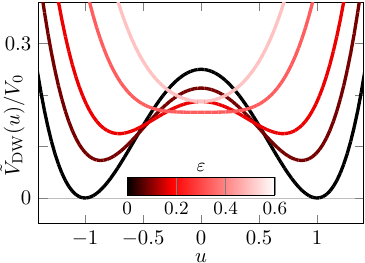}
\caption{Effective obstacle in the presence of noise: Noise-averaged penalty $\tilde V_\text{DW}(u)$ for increasing noise level $\eps$ [\Eqnref{epsdef}, color bar]. Thermal averaging progressively smears the double
    well [\Eqnref{dwell-main}] into a single well.}
    \figlab{potential}
\end{figure}

\paragraph*{Mean work as a special case.} A prominent physical example of the cost \Eqnref{cost} is the
mean stochastic work performed on the particle by the moving trap~\cite{Mei26b}, given by~\cite{Sek10}
\algn{\eqnlab{Wstoch}
    W_{t_f} = \int_0^{t_f}\!\!\ed t\,\kappa\dot\lambda(t)[\lambda(t)-x(t)]
    + V(x(t_f))\,.
}
Taking the noise average $\langle\ldots\rangle$ and integrating by parts, the mean of the path-dependent
contribution becomes
\algn{\eqnlab{Wmean_partial}
    &\left\langle\int_0^{t_f}\!\!\ed t\,\kappa\dot\lambda(\lambda-x)\right\rangle
    = \kappa\left[\lambda\left(\frac{\lambda}2 - u\right)\right]_0^{t_f}
   + \kappa\int_0^{t_f}\!\!\ed t\,\lambda\dot u\nn\\
    &= \kappa\left[\lambda\left(\frac{\lambda}2 - u\right)\right]_0^{t_f}
    + \frac{\kappa}{\tau_\text{p}}\int_0^{t_f}\!\!\ed t\,\lambda
    (\lambda-u)\,.
}
This can be written in the quadratic form
$\int_0^{t_f}\ed t\,\vvec^{\sf T}\Cmat_W\vvec + F_W(\vvec_f;\vvec_0)$ with
\algn{\eqnlab{CmatW}
    \Cmat_W = \frac{\kappa}{2\tau_\text{p}}\pmat{0 & -1 \\ -1 & 2}\,,
    \qquad
    \Bmat{f,0}_W = \frac{\kappa}{2}\pmat{0 & -1 \\ -1 & 1}\,.
}
We note that $C_{11} = 0$ for the mean work: the running cost does not
directly penalize the mean position, only the control velocity and the
mismatch between trap and particle. As we show in
\Secref{hamiltonian}, the mean-work problem falls in the \textit{parabolic
class}.

\paragraph*{Quadratic control cost.} A second physically motivated cost functional is the mean energy dissipated
against the internal friction of the control mechanism, which is
proportional to the integrated squared control amplitude~\cite{Bec21},
\algn{\eqnlab{Cquad}
    \ms{C}^\text{ctrl}_{t_f} = c\int_0^{t_f}\!\!\ed t\,\lambda^2(t) + \tilde V(u_f)\,,
}
for some constant $c > 0$. This cost corresponds to choosing
\algn{\eqnlab{CmatCtrl}
    \Cmat_\text{ctrl} = \pmat{0 & 0 \\ 0 & c}\,,\qquad \Bmat{f,0}_\text{ctrl} = 0\,.
}
For this cost combined with the harmonic-trap dynamics,
we will show in \Secref{hamiltonian} that the control problem falls in the
\textit{hyperbolic class} with a direct connection to the
finite-time dynamical phase transition in the relaxation into a harmonic
potential (studied in \Secref{mapping} and \Secref{experiments}).

In \Secref{canonical}, we construct an example similar to \Eqnref{Cquad} that falls into the elliptic class.

% =========================================================
\section{Hamiltonian Formulation and Canonical Transformations}
\seclab{hamiltonian}
% =========================================================

% -------------------------------------------------------
\subsection{Reduction to Hamiltonian form}
\seclab{pontryagin}
% -------------------------------------------------------

We now derive the optimal control protocol $\lambda^*(t)$ by applying the
Pontryagin minimum principle~\cite{Fle75,Bec21} to the cost functional
\Eqnref{cost} subject to the mean dynamics \Eqnref{meandyn}. The derivation
proceeds in three steps: we introduce a Lagrange multiplier to enforce the
mean dynamics~\eqnref{meandyn}, eliminate the control $\lambda(t)$ at each time $t$
by pointwise minimization, and identify the resulting structure as a quadratic
control Hamiltonian.

\paragraph*{Incorporating the dynamical constraint.}
We enforce the mean dynamics~\eqnref{meandyn} by introducing a Lagrange
multiplier $\mu(t)$ and forming the augmented cost
\begin{multline}\eqnlab{augmented}
    \ms{C}_{t_f}[u(t),\lambda(t),\mu(t)] =  F(\vvec_f;\vvec_0) + \tilde V(u_f)
    \\+
    \int_0^{t_f}\!\!\ed t\,
    \Big\{\vvec^{\sf T}\Cmat\vvec + \mu(t)\big[\dot u(t) + \tau_\text{p}^{-1}[u(t) - \lambda(t)]\big]\Big\}
    \,.
\end{multline}
The Lagrange multiplier $\mu(t)$ plays the role of the conjugate momentum
to the generalized coordinate $u(t)$ in the Hamiltonian formulation that
emerges below.

\paragraph*{Gauge freedom of the cost.}
Before minimizing, we note a ``gauge freedom'' of the cost functional that we
will use to simplify the control Hamiltonian below. Adding and subtracting the total
time derivative $\frac{\alpha}{2}\frac{\ed}{\ed t}u^2(t)$ to the integrand of
\Eqnref{augmented} leaves the optimal-control problem invariant but changes $\Cmat$
and $\Bmat{f,0}$ according to
\algn{\eqnlab{Cshift}
    \Cmat \;\to\; \Cmat + \frac{\alpha}{\tau_\text{p}}
    \pmat{1 & -\tfrac{1}{2} \\ -\tfrac{1}{2} & 0}\,,
    \quad
    \Bmat{f,0} \;\to\; \Bmat{f,0} + \frac{\alpha}{2}
    \pmat{1 & 0 \\ 0 & 0}\,,
}
for any real parameter $\alpha$. We carry $\alpha$ along in the following and fix it conveniently to diagonalize the Hamiltonian matrix of the control Hamiltonian in
\Secref{canonical}.

\paragraph*{Pointwise minimization over the control.}
The Pontryagin minimum principle asserts that for fixed $u(t)$ and $\mu(t)$,
the optimal control $\lambda^*(t)$ minimizes the integrand of \Eqnref{augmented}
at each time $t\in(0,t_f)$, and similarly minimizes the boundary functional
$F$ at the endpoints. Taking the variation of \Eqnref{augmented} with
respect to $\lambda(t)$ and setting it to zero, 
\algn{
	\frac{\delta \ms{L}}{\delta\lambda(t)}\bigg|_{\lambda=\lambda^*}\!\!\!\!= 0\,,
}
where $\ms{L}$ denotes the integrand in \Eqnref{augmented}, gives
\algn{\eqnlab{lambdaopt}
    \lambda^*(t) = \frac{\mu(t) +(\alpha- 2C_{12}\tau_\text{p})u(t)}{2C_{22}\tau_\text{p}}\,.
}

Minimizing the boundary terms $F$ over the endpoint values $\lambda_{f,0}$ yields
\algn{\eqnlab{lambdabc}
    \lambda^*_{f,0} = (1-\Delta_{f,0})u_{f,0}\,,\qquad \Delta = \frac{B^{f,0}_{12}+B^{f,0}_{22}}{B^{f,0}_{22}}\,,
}
provided $B^{f,0}_{22}\neq 0$. The possible mismatch between bulk limits $\lambda^*(0^+)$ and $\lambda^*(t_f^-)$ in \Eqnref{lambdaopt} with $\lambda^*_0$ and $\lambda^*_f$ in \Eqnref{lambdabc}, respectively, enables discontinuous jumps of the optimal protocol $\lambda^*(t)$ at the initial and endpoints, a common feature of such control problems~\cite{Sch07,Moh25}.

 The boundary term $F$ is quadratic in
$\lambda_{f,0}$, so three cases in \Eqnref{lambdabc} must be distinguished. For $B^{f,0}_{22}>0$, $F$ is
strictly convex and \Eqnref{lambdabc} is its unique minimizer. In this case, boundary jumps are generic. If
$B^{f,0}_{22} = 0$ (or $B^{f,0}_{22} < 0$, $B^{f,0}_{12}\neq 0$), the boundary terms are unbounded below, which renders the minimization over $\lambda_{f,0}$ ill-posed. For vanishing  $B^{f,0}_{12} = B^{f,0}_{22} = 0$, the boundary term is
independent of $\lambda_{f,0}$ altogether, so endpoint control is unconstrained and can be chosen to match the limits of the bulk solution~\eqnref{lambdaopt}, so that $\lambda^*(t)$ has no jumps.

The coefficients $\Delta_{f,0}$ are a property of the boundary matrices
alone and are therefore an independent input to the problem. For the mean work~\eqnref{CmatW} one has $B^{f,0}_{11}=0$ and
$B^{f,0}_{22}=-B^{f,0}_{12}=\kappa/2$, so $\Delta_{f,0}=0$ and $\lambda^*_{f,0}=u_{f,0}$. Consequently, the
trap coincides with the mean particle position at both ends, ensuring that no excess potential energy is stored.

\paragraph*{Quadratic control Hamiltonian.}
Substituting the optimal control~\eqnref{lambdaopt} back into the augmented
cost~\eqnref{augmented}, using the shift~\eqnref{Cshift}, and collecting
terms, we arrive at an expression for $\ms{C}_{t_f}$ that depends only on $u(t)$ and $\mu(t)$:
\begin{multline}\eqnlab{Hform}
    \ms{C}_{t_f}[u(t),\lambda(t)] = \int_0^{t_f}\!\!\ed t\,\Big\{\mu(t)\dot u(t) - \ms{H}[u(t),\mu(t)]\Big\}
    \\+ G(u_0,u_f) + \tilde V(u_f)\,,
\end{multline}
where the \emph{control Hamiltonian} $\ms{H}$ is the quadratic form
\algn{\eqnlab{Hquad}
    \ms{H}(u,\mu) = \frac{1}{2}\wvec^{\sf T}\,\Hmat\,\wvec\,,
    \qquad \wvec = \pmat{u(t)\\\mu(t)}\,,
}
and $G(u_0,u_f)$ collects the boundary terms. The Hamiltonian matrix $\Hmat$
has components
\sbeqs{\eqnlab{Hmatcomps}
\algn{
    H_{11} &= -\frac{2}{C_{22}}\left[\left(C_{11}+\frac{\alpha}{\tau_\text{p}}\right)C_{22}
    - \left(C_{12} - \frac{\alpha}{2\tau_\text{p}}\right)^2\right]\,,
    \eqnlab{H11}\\
    H_{12} &= H_{21} = -\frac{1}{C_{22}}\left(C_{12} + C_{22}
    - \frac{\alpha}{2\tau_\text{p}}\right)\,,
    \eqnlab{H12}\\
    H_{22} &= \frac{1}{2C_{22}\tau_\text{p}^2}\,,
    \eqnlab{H22}
}
}
and the simplified boundary terms read
\algn{\eqnlab{Gbound}
    G(u_0,u_f) = \ms{B}_f\,u_f^2 - \ms{B}_0\,u_0^2\,,
}
with scalar coefficients
\algn{\eqnlab{Bscalar}
    \ms{B}_{f,0} = B^{f,0}_{11} + \frac{\alpha}{2}
    - \frac{(B^{f,0}_{12})^2}{B^{f,0}_{22}}\,,
}
whenever $B^{f,0}_{22}\neq 0$. In the unconstrained case
$B^{f,0}_{12}=B^{f,0}_{22}=0$, we instead have
$\ms{B}_{f,0} = B^{f,0}_{11} + \alpha/2$.

In the Hamiltonian formulation, $u(t)$ plays the role of a generalized
coordinate and $\mu(t)$ of its conjugate momentum. The pair $\wvec(t) = [u(t),\mu(t)]^{\sf T}$
evolves in a two-dimensional phase space.

\paragraph*{Hamilton equations and boundary conditions.}
Requiring stationarity of \Eqnref{Hform} with respect to variations of
$u(t)$ and $\mu(t)$ at fixed endpoints $u_0$ and $u_f$ yields the
Hamilton equations of motion
\algn{\eqnlab{Hamilton}
    \dot{\wvec^*}(t) = \Jmat\Hmat\,\wvec^*(t)\,,
    \qquad
    \Jmat = \pmat{0 & 1 \\ -1 & 0}\,,
}
for the optimal solution $\wvec^*(t)$, where $\Jmat$ is the standard symplectic matrix~\cite{Arn01}. The general solution is
\algn{\eqnlab{Hamsol}
    \wvec^*(t) = \mbb{M}_t\,\wvec^*(0)\,,
}
where $\mbb{M}(t)$ denotes the symplectic propagator
\algn{\eqnlab{Mmat}
	\mbb{M}_t = \ee^{\Jmat\Hmat t}\,.
}
From \Eqnref{Hamsol}, the values $\mu_0=\mu(0)$ and $\mu_f=\mu(t_f)$ are connected to the boundary
conditions $u_0$ and $u_f$. Setting $u(0) = u_0$ and $u(t_f) = u_f$ in \Eqnref{Hamsol}
gives a linear system for $\mu(0)$ and $\mu(t)$ in terms of $u_0$ and
$u_f$:
\algn{\eqnlab{bcmat}
    \pmat{u^*(t)\\\mu^*(t)} = \mbb{M}_{t}\pmat{u_0\\\mu^*_0}\,,
}
which after solving for $[\mu^*_0,\mu^*(t)]^{\sf T}$ reads
\algn{\eqnlab{bcmat2}
    \pmat{\mu_0^*\\\mu^*(t)} = \mbb{\tilde M}_{t}\pmat{u_0\\ u(t)}\,,\quad \mbb{\tilde M}_{t} = M^{-1}_{12}(t)\pmat{-M_{11}(t) & 1\\ -1 & M_{22}(t)}\,.
}
The block $M_{12}(t)$ is invertible for all $t\in(0,t_f)$, so that $\mu_0^*$ and $\mu^*(t)$ are uniquely fixed by the boundary positions. This fails only in the elliptic class (defined in \Secref{canonical}) at $t=\pi\tau_\text{c}$, where $M_{12}=0$ and the elliptic problem becomes unstable, as we discuss in \Secref{transition}.
Upon setting $t=t_f$ in \Eqnref{bcmat2}, we obtain $\mu^*_0$ and $\mu^*_f$ from $u_0$ and $u_f$. This allows us to eliminate $\mu^*_0$ from \Eqnref{bcmat}, leading us to
\sbeqs{\eqnlab{Nmatall}
\algn{\eqnlab{Nmat}
    \pmat{u^*(t)\\\mu^*(t)} = \mbb{N}_{t}\pmat{u_0\\u_f}\,,
}
with the matrix
\algn{\eqnlab{Nmat2}
	\mbb{N}_{t} 	&= M_{12}^{-1}(t_f)\mbb{M}_{t}\pmat{M_{12}(t_f)	& 0 \\ -M_{11}(t_f)	& 1}\,,\\
				&= M^{-1}_{12}(t_f)\pmat{M_{12}(t_f)M_{11}(t)-M_{11}(t_f)M_{12}(t)	&	M_{12}(t)\\	M_{12}(t_f)M_{21}(t) - M_{11}(t_f)M_{22}(t)		&	M_{22}(t)}\,.\nn
}
}
Once $u^*(t)$ and $\mu^*(t)$ are known, the optimal protocol
$\lambda^*(t)$ and its boundary jumps follow from \Eqnref{lambdaopt}.

Recall that in our setup, $u_0$ is fixed while $u_f$ is free and determined by a
further minimization of $\ms{C}_{t_f}$ over $u_f$, as described in
\Secref{transition}. It is therefore natural to choose the boundary matrices
so that $\ms{B}_0 = 0$, which eliminates the boundary penalty on the
initial condition. Furthermore, the remaining boundary terms $\ms{B}_f u_f^2$ can
be absorbed into the terminal cost $\tilde V(u_f)$, so we set $\ms{B}_f = 0$
as well~\footnote{Note that for the mean work~\eqnref{CmatW} with the
diagonalizing choice $\alpha^*$ in \Eqnref{alphastar} below, $\ms{B}_{f,0}=0$ identically. The mean-work problem
therefore requires no boundary terms to be discarded or absorbed.}.

By an integration by parts, the cost function $\ms{C}_{t_f}$ along the optimal solution $\wvec^*(t)$ is written as
\begin{multline}\eqnlab{Costopt}
	\ms{C}_{t_f}[u^*(t),\lambda^*(t)] = \int_0^{t_f}\!\!\ed t\,\Big\{\frac12{\dot\wvec}^{*{\sf T}}(t)\mbb{J}\wvec^*(t) - \ms{H}[\wvec^*(t)]\Big\}
    \\+ \frac{1}{2}(u_f\mu_f - u_0\mu_0) + \tilde V(u_f)\,.
\end{multline}
Now, using \Eqnref{Hamilton}, $\mbb{H}^{\sf T} = \mbb{H}$, and $\Jmat^{\sf T}\Jmat = \mbb{1}$, we observe that
\algn{
	\frac12{\dot\wvec}^{*{\sf T}}\!(t)\mbb{J}\wvec^*(t) = \frac12{\wvec}^{*\sf T}\!(t)\mbb{H}\mbb{J}^{\sf T}\mbb{J}\wvec^*(t) = \ms{H}[\wvec^*(t)]\,,
}
which cancels with the second term in \Eqnref{Costopt}, so the cost evaluated along the optimal trajectory
simplifies to
\algn{
    \ms{C}_{t_f}(u_f,u_0) = \frac{1}{2}(u_f\mu^*_f - u_0\mu^*_0) + \tilde V(u_f)\,.
}
Expressing $\mu^*_0$ and $\mu^*_f$ in terms of $u_0$ and $u_f$
via \Eqnref{bcmat2} at $t=t_f$, we obtain
\sbeqs{\eqnlab{Copt}
\algn{
    \ms{C}_{t_f}(u_f,u_0) = \frac{1}{2}\ve u^{\sf T} \mbb{D}_{t_f} \ve u + \tilde V(u_f)\,,
}
where
\algn{
	\ve u = \pmat{u_0\\u_f}\,,\quad \mbb{D}_{t_f} = M^{-1}_{12}(t_f)\pmat{M_{11}(t_f)& -1\\ -1 & M_{22}(t_f)}\,.
}
}
Equations~\eqnref{Copt} are the central object of the analysis:
minimizing $\ms{C}_{t_f}$ over $u_f$ yields the optimal cost and the optimal final
position $u_f^*$, from which the full optimal protocol $\lambda^*(t)$ is
recovered via \Eqnref{lambdaopt} and \Eqnref{Hamsol}.

\paragraph*{Structure of the problem.}
The dynamics of the canonical vector $\wvec(t)$ are equivariant under
symplectic (canonical) transformations $\wvec\to\wvec' = \Smat\wvec$
generated by $2\times 2$ matrices $\Smat$ with unit determinant satisfying~\cite{Arn01}
\algn{\eqnlab{symplectic}
    \Smat\,\Jmat\,\Smat^{\sf T} = \Jmat\,,
}
which fixes $\det\Smat=1$. Such transformations leave $\Jmat$ invariant and map solutions of the
Hamilton equations to solutions. They transform $\Hmat$ according to
\algn{\eqnlab{strafo}
	\Hmat\to\Smat^{\sf T}\Hmat\Smat\,,
}
while leaving the cost invariant up to boundary terms absorbed into $G$; cf.~\Eqnref{Hform}. The gauge transformation~\eqnref{Cshift} is precisely a canonical transformation of this type with
\algn{\eqnlab{Smat}
    \Smat = \pmat{1 & 0 \\ \alpha & 1}\,,
}
as can be verified directly from applying~\Eqnref{strafo} with~\Eqnref{Smat} to $\mbb{H}$ and comparing with~\Eqnref{Hmatcomps}. Hence in the Hamiltonian formalism applied here, the ``gauge transformation'' introduced above corresponds to a specific class of canonical transformation that leaves $u$ invariant, $u'=u$. 

These observations are key to our classification of all control problems of the type described in \Secref{cost} into three canonical
equivalence classes, which we develop next.

% -------------------------------------------------------
\subsection{Three equivalence classes}
\seclab{canonical}
% -------------------------------------------------------

The Hamiltonian matrix $\Hmat$ derived in \Secref{pontryagin} is in general
not diagonal, and its off-diagonal elements $H_{12}$ depend on the free
parameter $\alpha$~\eqnref{Cshift}. We now choose $\alpha$ to
diagonalize $\Hmat$, and then use a further symplectic rescaling to bring it
into one of three canonical normal forms. The resulting classification
exhausts all possible control problems of the form~\eqnref{cost}
up to canonical equivalence.

\paragraph*{Diagonalization by choice of $\alpha$.}
Setting $H_{12} = 0$ in \Eqnref{H12} gives the condition
\algn{\eqnlab{alphastar}
    \alpha^* = 2(C_{12} + C_{22})\tau_\text{p}\,.
}
With this choice, the Hamiltonian matrix becomes diagonal,
\algn{\eqnlab{Hdiag}
    \Hmat = \pmat{\zeta & 0 \\ 0 & \xi}\,,
}
with eigenvalues
\algn{\eqnlab{zetaxi}
    \zeta = -2(C_{11} + 2C_{12} + C_{22})\,,
    \qquad
    \xi = \frac{1}{2C_{22}\tau_\text{p}^2}\,.
}
and determinant
\algn{\eqnlab{detH}
    \det\Hmat = \zeta\xi\,.
}

The parameter $\xi$ is set by the particle relaxation timescale $\tau_\text{p}$ and the cost $C_{22}$ of the control and determines the \textit{overall energy scale} of the Hamiltonian. Since
$C_{22} > 0$ by assumption, $\xi$ is
always positive. 
The sign of $\det\Hmat$ is thus entirely determined by the sign of
$\zeta$, which depends on the cost matrix $\Cmat$ and is not fixed
a priori. Physically, $\zeta$ determines how the cost responds
to simultaneous displacements of $u$ and $\lambda$ from their zero-control values.
For instance, the mean-work cost~\eqnref{CmatW} depends only on
the trap--particle mismatch $\lambda-u$ and the control velocity $\dot\lambda$, and is
therefore invariant under a simultaneous shift $u\to u+a$, $\lambda\to\lambda+a$. This is a consequence of the homogeneity of space away from the obstacle for $\zeta=0$. By contrast, when $\zeta < 0$ such mutual $\lambda$-$u$ displacements are penalized by the running cost and when $\zeta > 0$ they are beneficial. As we show below, the sign of $\zeta$ turns out to classify all control problems of the form~\eqnref{cost} into three canonical equivalence classes, each with a distinct normal form for $\Hmat$ and a qualitatively different solution structure. 

With $\alpha^*$ fixed, the optimal protocol follows from \Eqnref{lambdaopt} evaluated along the optimal trajectory,
\algn{\eqnlab{lambdastar}
    \lambda^*(t) = u^*(t) + \tau_\text{p}\xi\,\mu^*(t)\,,
}
and its boundary jumps at $t=0$ and $t=t_f$ read
\sbeqs{\eqnlab{Deltalam}
\algn{
    \Delta\lambda_{0} &= \lambda^*(0^+) - (1-\Delta_0)u_0 = \tau_\text{p}\xi\,\mu^*(0) + \Delta_0u_0\,,\\
    \Delta\lambda_{f} &=(1-\Delta_f)u_f - \lambda^*(t_f^-) = -\tau_\text{p}\xi\,\mu^*(t_f)-\Delta_fu_f\,,
}
}
where we used \Eqnref{lambdabc} and the continuity of $\mu^*(t)$.

\paragraph*{Normal forms via symplectic rescaling.}
After diagonalization, the Hamilton equations \eqnref{Hamilton} take the
decoupled form $\dot u = \xi\mu$ and $\dot\mu = -\zeta u$. In an additional step,
we now show how to bring the Hamiltonians $\ms{H}$ into a standard normal form,
which allows us to easily classify them. To this end, for $\zeta \neq 0$, we apply the additional
symplectic rescaling
\algn{\eqnlab{Srescale}
    \Smat_\text{res} = \text{diag}\!\left(|\xi/\zeta|^{1/4}, |\zeta/\xi|^{1/4}\right)\,,
}
which satisfies \Eqnref{symplectic} and maps
$\Hmat\to\Smat_\text{res}^{\sf T}\Hmat\Smat_\text{res}$. The
transformed Hamiltonian matrix takes the normal form
\algn{\eqnlab{Hnormal}
    \Hmat'_\eta = \frac{1}{\tau_\text{c}}\pmat{\eta & 0 \\ 0 & 1}\,,
}
where $\eta = \text{sgn}(\zeta)$ and the \emph{control timescale}
\algn{\eqnlab{tauc}
    \tau_\text{c} = |\zeta\xi|^{-1/2}
}
sets the natural time unit for the transition dynamics. In the transformed
coordinates $\ve{w}'(t) = \Smat_\text{res}\wvec(t)$, the control Hamiltonian
reads
\algn{\eqnlab{Hnormalquad}
    \ms{H}' = \frac{1}{2\tau_\text{c}}\left(\mu'^2 + \eta\, u'^2\right)\,,
}
which describes a harmonic oscillator ($\eta = +1$) or a particle in an
inverted parabola ($\eta = -1$). The case $\eta=\zeta = 0$ is treated separately by the normal form
\algn{\eqnlab{H0normal}
    \Hmat_0 = \frac{1}{\kappa\tau_0}\pmat{0 & 0 \\ 0 & 1}\,,
}
where $\tau_0 = 1/(\kappa\xi)$ is the corresponding control timescale.

Although the rescaling by $\Smat_\text{res}$ is convenient for the sake
of classifying the control Hamiltonian $\mbb{H}$, it bears the complication that
it transforms the position variable, $u' = |\xi/\zeta|^{1/4}u$, and therefore also
deforms the terminal cost, $\tilde V(u_f)=\tilde V(|\zeta/\xi|^{1/4}u_f')$. In other words,
$\mbb{S}_\text{res}$ stretches or squeezes the obstacle.

In fact, it is straightforward to check that the gauge transformation by $\Smat$ in \Eqnref{Smat} is the \emph{only}
canonical transformation in $d=1$ that leaves $u$ invariant. We therefore use \Eqnref{Srescale} solely
to \emph{classify} the Hamiltonians, for which the terminal cost is
irrelevant, and work with the diagonal form \Eqnref{Hdiag} whenever an
actual control problem is solved or compared with another one.

\paragraph*{Equivalence classes.}
The classification by $\eta=\text{sgn}(\zeta)=\text{sgn}(\det\Hmat)$ divides all linearized optimal
control problems of the form~\eqnref{augmented} into three
canonical equivalence classes, as advertised in~\Figref{overview}(b). Each class comprises a distinct normal form for $\Hmat$
and a qualitatively different solution structure:

\begin{itemize}
    \item \textit{Parabolic class} ($\zeta = 0$, $\det\Hmat = 0$): the
    Hamiltonian is $\ms{H}_0 = \mu^2/(2\kappa\tau_0)$,
    describing purely ballistic motion in the $u$-direction, i.e., corresponding to a free particle in classical mechanics. The Hamilton equations
    give $\dot u^* = \xi\mu^* = \text{const}$, so the optimal mean trajectory
    $u^*(t)$ is linear in time. The mean-work problem
    for the harmonic trap studied in \cite{Mei26b} falls into the parabolic class. 

    \item \textit{Hyperbolic class} ($\zeta < 0$, $\det\Hmat < 0$): the
    Hamiltonian is $\ms{H}'\propto\mu'^2 - u'^2$, and the Hamilton
    equations are those of a particle in an inverted quadratic potential. The
    solutions are combinations of $\sinh$ and $\cosh$, and the cost
    function decays monotonically with protocol duration. A physical
    realization is the harmonic-trap dynamics with the quadratic control
    cost \Eqnref{Cquad}, as we verify in \Secref{hamclass}.

    \item \textit{Elliptic class} ($\zeta > 0$, $\det\Hmat > 0$): the
    Hamiltonian is $\ms{H}'\propto\mu'^2 + u'^2$, and the Hamilton
    equations are those of a harmonic oscillator. Solutions are
    combinations of $\sin$ and $\cos$. 
    
\end{itemize}
\paragraph*{Examples:}
We now look at three example control problems for a system with the linear dynamics~\eqnref{meandyn}.

First, if the control goal is the mean-work cost \Eqnref{CmatW}, direct substitution into \Eqnref{zetaxi} gives
\algn{
    \zeta_W = -2\left(\frac{\kappa}{2\tau_\text{p}}\right)(0 - 1 + 1) = 0\,,
    \qquad
    \xi_W = \frac{1}{2\tau_\text{p}\kappa}\,,
}
confirming that work cost functions correspond to the parabolic class ($\zeta=0$), with control timescale
$\tau_0 = 1/(\kappa\xi_W) = 2\tau_\text{p}$. This emphasizes the important role of the parabolic class, which represents not just a degenerate special case but encompasses important physical cost functions, due to its link to spatial homogeneity.
For the mean work specifically, both the cost~\eqnref{CmatW} and the linearized dynamics~\eqnref{meandyn} depend only on the control velocity $\dot\lambda$ and on the trap--particle mismatch $\lambda - u$, making the invariance explicit. The obstacle is the only ingredient breaking this invariance, doing so solely through the terminal cost $\tilde V$. Consequently, the classification involves the control \emph{dynamics} and is independent of the obstacle. In particular, it applies equally to the classical minimum-work problem of~\cite{Sch07}, where no terminal cost is present.

Second, if the control goal is to minimize the quadratic control cost \Eqnref{CmatCtrl} with $c>0$, substitution then gives
\algn{\eqnlab{zetahyp}
    \zeta_\text{ctrl} = -2(0 + 0 + c) = -2c < 0\,,
    \qquad
    \xi_\text{ctrl} = \frac{1}{2c\tau_\text{p}^2}\,,
}
confirming that such a control problem belongs to the hyperbolic class. Because this cost contains no $\dot\lambda$, it generates
no boundary term, $\Bmat{f,0}_\text{ctrl}=0$, and hence corresponds to the unconstrained
case discussed below \Eqnref{lambdabc}: the endpoint protocol values are free and the
optimal protocol can be chosen to have no jumps (\Secref{hamclass}). The control timescale is
$\tau_\text{c} = |\zeta_\text{ctrl}\xi_\text{ctrl}|^{-1/2} = \tau_\text{p}$,
so in this case $\tau_\text{c}$ coincides with the dynamical
relaxation time $\tau_\text{p}$, independently of the cost coefficient $c$.

Finally, a cost that \emph{rewards} displacement from the
zero-control position realizes the elliptic class. An example of this is obtained by augmenting the quadratic control cost
\Eqnref{Cquad} by a running reward for finite $u(t)$,
\algn{\eqnlab{Cavoid}
    \ms{C}^\text{av}_{t_f} = c\int_0^{t_f}\!\!\ed t\,
    \left[\lambda^2(t) - \wp\,u^2(t)\right] + \tilde V(u_f)\,,
}
with the parameter $\wp$. With this control, we have $C_{11}=-c\wp$, $C_{12}=0$ and $C_{22}=c$.
Compared with \Eqnref{Cquad}, \Eqnref{Cavoid} incorporates the additional control objective to avoid the region around $u=0$: the system is
rewarded for staying away from $u=0$ while paying for
the actuation needed to do so. Substituting into \Eqnref{zetaxi} gives
\algn{\eqnlab{zetaavoid}
    \zeta_\text{av} = 2c(\wp-1)\,,\qquad
    \xi_\text{av} = \frac{1}{2c\tau_\text{p}^2}\,,
}
which shows that $\zeta_\text{av}$ can have either sign and all three classes are reached by tuning $\wp$: The problem is
hyperbolic for $\wp<1$, parabolic at $\wp=1$, and elliptic for $\wp>1$, with control
timescale $\tau_\text{c} = \tau_\text{p}/\sqrt{\wp-1}$ in the elliptic case. For $\wp=1$, the cost is invariant under
simultaneous translations of $\lambda$ and $u$, similar to the mean work, up to a boundary term that is linear in $u_{f,0}$. The quadratic control cost~\eqnref{Cquad} is recovered for $\wp=0$.

\paragraph*{Relation to classical normal-form theory.}
Reducing a quadratic Hamiltonian to a normal form under linear canonical
transformations is the problem solved in general dimension by Williamson~\cite{Wil36,Arn01}. Our identification of $\zeta$ (the sign of $\det\Hmat$) with the response of the cost to simultaneous displacements of mean position and control
endows the three classes in $d=1$ with a physical meaning in the context of optimal control and ties the parabolic class to the homogeneity of space,
thereby explaining why mean-work optimization is marginal, i.e., $\zeta=0$.

% -------------------------------------------------------
\subsection{Cost functions and optimal protocols}
\seclab{hamclass}
% -------------------------------------------------------

With the normal forms \Eqnref{Hnormal} and \Eqnref{H0normal} established,
we now solve the Hamilton equations \Eqnref{Hamilton} for each class,
evaluate the cost \Eqnref{Copt} along the optimal trajectory, and derive
the corresponding optimal protocol $\lambda^*(t)$ via \Eqnref{lambdaopt}.

\paragraph*{Solution of the Hamilton equations.}
In each class, the symplectic propagator $\mbb{M}_t = \ee^{\Jmat\Hmat t}$ is computed from
the diagonal form~\eqnref{Hdiag} of $\Hmat$. We work directly with the diagonalized Hamiltonian and omit the further symplectic rescaling \Eqnref{Srescale}, which is not needed to solve the Hamilton equations and would deform the terminal cost, as noted in \Secref{canonical}. We now discuss all three classes.

\paragraph*{Parabolic class ($\zeta = 0$).}
For the parabolic class, the symplectic propagator~\eqnref{Mmat} and the matrix $\mbb{N}_t$~\eqnref{Nmat2} read 
\algn{\eqnlab{MN0}
	\mbb{M}_t^0 = \pmat{1 & \xi t\\ 0 & 1}\,, \quad  	\mbb{N}_t^0 = t_f^{-1}\pmat{t_f- t & t\\ -\xi^{-1} & \xi^{-1}}\,,
}
so we find from \Eqnref{Nmat}:
\algn{\eqnlab{solpara}
    u^*(t) = (u_f - u_0)\frac{t}{t_f} + u_0\,,\quad     \mu^*(t) = \frac{u_f - u_0}{\xi t_f}\,.
}
We thus find that $u^*(t)$ interpolates linearly between $u_0$ and $u_f$, while $\mu^*(t)$ remains constant. This is expected, considering that in the Hamiltonian formalism, $u^*(t)$ and $\mu^*(t)$ correspond to the position and momentum, respectively, of a free particle. The linear mean position $u^*(t)$ and the trajectory of the conjugate parameter $\mu^*(t)$ for the parabolic class are shown as the red lines in \Figref{protocols}(a) and \figref{protocols}(c), respectively.

Evaluating \Eqsref{Copt} with \eqnref{MN0} gives the cost
\algn{\eqnlab{C0}
    \ms{C}^0_{t_f}(u_f, u_0) = \frac12\frac{(u_f - u_0)^2}{\xi t_f} + \tilde V(u_f)\,.
}
For the mean-work cost with the harmonic trap, where $\xi_W = 1/(2\kappa\tau_\text{p})=1/(2\gamma)$,
\Eqnref{C0} reduces to the result of the companion letter~\cite{Mei26b}.
The first term is the dissipative cost of transporting the mean position
from $u_0$ to $u_f$ in time $t_f$. It diverges as $t_f\to 0$ and vanishes
as $t_f\to\infty$, reflecting the trade-off between speed and dissipation: fast transport (small $t_f$) drives the particle against friction more rapidly and dissipates more work, whereas slow transport approaches the quasistatic limit at vanishing dissipative cost.

The optimal protocol follows from \Eqnref{lambdastar} and \Eqnref{solpara}:
\algn{\eqnlab{lam0}
    \lambda^*(t) = \css{
        (1-\Delta_0)u_0\,, & t = 0\,,\\[4pt]
        \dfrac{u_f - u_0}{t_f}\left(t + \tau_\text{p}\right) + u_0\,, & 0 < t < t_f\,,\\[8pt]
        (1-\Delta_f)u_f\,, & t = t_f\,.
    }
}
\begin{figure*}
	\includegraphics{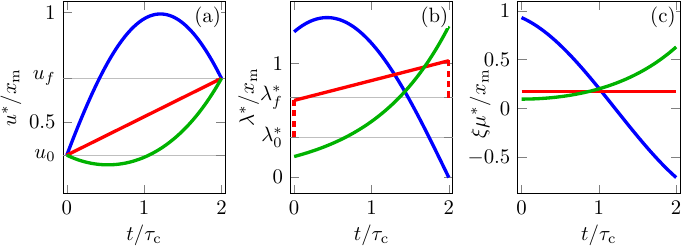}
	\caption{Examples of optimal trajectories $(u^*, \lambda^*,\mu^*)$ for the different equivalence classes, $\eta=\text{sign}(\zeta)=-1$ (hyperbolic class, green), $\eta=0$ (parabolic, red), and $\eta=1$ (elliptic, blue), for identical fixed endpoints $u_0= 0.35\,x_\text{m}$ and $u_f = 0.7\,x_\text{m}$, as a function of time $t$ for $t_f = 2\tau_\text{c}$. (a) Optimal mean particle trajectories $u^*(t)$. (b) Optimal protocols $\lambda^*(t)$. The parabolic, mean-work case has jumps at the start and endpoints (dashed lines) for $\Delta_{f,0}=0$ in \Eqnref{lambdabc}, while the other cases were chosen continuous at the endpoints. (c) Optimal conjugate momenta $\mu^*(t)$.}
	\figlab{protocols}
\end{figure*}
The protocol is piecewise linear with discontinuous jumps
$\Delta\lambda_{f,0} = \mp(u_f - u_0)\tau_\text{p}/t_f\mp\Delta_{f,0}u_{f,0}$ at both endpoints. For the mean work with $\Delta_{f,0}=0$, these jumps are of equal height and opposite direction and establish the trap--particle mismatch that
sustains a constant mean velocity throughout the protocol, as first
identified for the harmonic-trap mean-work problem in~\cite{Sch07}. The optimal protocol $\lambda^*(t)$ for the parabolic class, including the jumps corresponding to $\Delta_{f,0}=0$, is shown as the red line in \Figref{protocols}(b).

Note also that up to the endpoint jumps of $\lambda^*(t)$, the bulk of the optimal solutions $\lambda^*(t)$ and $u^*(t)$ in the parabolic class ($\zeta=0$) are universal across the entire class, when the endpoints $u_0$ and $u_f$ are held fixed. By contrast, the conjugate momentum $\mu^*(t)$ and the cost $\ms{C}^0_{t_f}$ depend on $\xi$ and are thus not universal across the class. 

\paragraph*{Hyperbolic class ($\zeta < 0$).}
In the hyperbolic class, the Hamilton equations take the form of a particle in an inverted
parabolic potential with solutions in the form of hyperbolic sines and cosines.
In this case, the matrix exponential $\mbb{M}^-_t = \ee^{\Jmat\Hmat^- t}$ and $\mbb{N}^-_t$ involve hyperbolic functions:
\algn{
	\mbb{M}^-_t &= \pmat{\cosh\left(\frac{t}{\tau_\text{c}}\right)	&	\tau_\text{c}\xi\sinh\left(\frac{t}{\tau_\text{c}}\right) \\ \frac1{\tau_\text{c}\xi}\sinh\left(\frac{t}{\tau_\text{c}}\right)	&	\cosh\left(\frac{t}{\tau_\text{c}}\right)}\,,\\
	\mbb{N}^-_t &= \frac{1}{\sinh\left(\frac{t_f}{\tau_\text{c}}\right)}\pmat{\sinh\left(\frac{t_f-t}{\tau_\text{c}}\right)	&	\sinh\left(\frac{t}{\tau_\text{c}}\right) \\ -\frac1{\tau_\text{c}\xi}\cosh\left(\frac{t_f-t}{\tau_\text{c}}\right)	&	\frac1{\tau_\text{c}\xi}\cosh\left(\frac{t}{\tau_\text{c}}\right)}\,.
}
We then find from \Eqnref{Nmat} the optimal solutions
\algn{\eqnlab{uhyp}
    u^*(t) = \frac1{\sinh\left(\frac{t_f}{\tau_\text{c}}\right)}\left[\sinh\left(\frac{t}{\tau_\text{c}}\right)u_f + \sinh\left(\frac{t_f-t}{\tau_\text{c}}\right)u_0\right]
    \,,
}
and
\algn{\eqnlab{solhyp}
    \mu^*(t) = \frac{1}{\tau_\text{c}\xi\sinh\left(\frac{t_f}{\tau_\text{c}}\right)}
    \left[\cosh\!\left(\frac{t}{\tau_\text{c}}\right)u_f
    - \cosh\!\left(\frac{t_f-t}{\tau_\text{c}}\right) u_0  
    \right]\,,
}
yielding through \Eqnref{lambdastar} the optimal protocol
\begin{multline}
	\lambda^*(t) = \frac1{\sinh\left(\frac{t_f}{\tau_\text{c}}\right)}\left\{\left[\sinh\left(\frac{t}{\tau_\text{c}}\right)+\frac{\tau_\text{p}}{\tau_\text{c}}\cosh\!\left(\frac{t}{\tau_\text{c}}\right)\right]u_f\right.\\
	\left. + \left[\sinh\left(\frac{t_f-t}{\tau_\text{c}}\right)-\frac{\tau_\text{p}}{\tau_\text{c}}\cosh\!\left(\frac{t_f-t}{\tau_\text{c}}\right)\right]u_0\right\}\,,
\end{multline}
for $0<t<t_f$, together with the endpoint values
$\lambda^*(0)=(1-\Delta_0)u_0$ and $\lambda^*(t_f)=(1-\Delta_f)u_f$ of
\Eqnref{lambdabc}. The corresponding jumps of the optimal protocol are
\algn{
	\Delta\lambda_{f,0} = -\frac{\tau_\text{p}}{\tau_\text{c}}\,
	\frac{\cosh\left(\frac{t_f}{\tau_\text{c}}\right)u_{f,0} - u_{f,0}}
	{\sinh\left(\frac{t_f}{\tau_\text{c}}\right)}
	\mp \Delta_{f,0}\,u_{f,0}\,,
}
which are in general non-zero and, in contrast to the parabolic case, of unequal magnitude at the two ends, even for $\Delta_{f,0}=0$. 

Note that the boundary matrices $\mbb{B}_{f,0}$ need not arise from an integration by parts of the cost, as they do for the mean work. In general, an independently specified boundary cost yields also $\Delta_{f,0}\neq 0$ and modifies the jumps accordingly. Only if the boundary cost $G(u_f,u_0)$ vanishes altogether, as for the quadratic control cost~\eqnref{Cquad} with
$\Bmat{f,0}_\text{ctrl}=0$, are the endpoint values unconstrained. In that
case we adopt the continuous assignment $\Delta\lambda_{f,0}=0$, which seems physically natural, because
any real control would incur a finite, albeit small, cost for an instantaneous displacement. 

The green lines in \Figsref{protocols}(a)--(c) show $u^*(t)$, $\lambda^*(t)$, and $\mu^*(t)$ for the hyperbolic class, where $\lambda^*(t)$ in \Figref{protocols}(b) is shown for the jump-free case $\Delta\lambda^*_{f,0}=0$.

Within the hyperbolic class, joint displacements of $u$ and $\lambda$ are penalized, which explains why both $u^*(t)$ and $\lambda^*(t)$ remain closer to the origin when compared with the corresponding trajectories for the parabolic class.

Finally, \Eqnref{Copt} gives the cost
\algn{\eqnlab{Cminus}
	\ms{C}^-_{t_f}(u_f, u_0) = \frac1{2\xi\tau_\text{c}}\,
    	\frac{(u_f^2 +u_0^2)\cosh\left(\frac{t_f}{\tau_\text{c}}\right)-2u_0u_f}{\sinh\left(\frac{t_f}{\tau_\text{c}}\right)}
    	+ \tilde V(u_f)\,.
}

The first term of \Eqnref{Cminus} is the dissipative cost for
the hyperbolic class. For $t\ll\tau_\text{c}$ it reduces to the parabolic result \Eqnref{C0}. This is expected because on durations short compared with $\tau_\text{c}$ the dynamics must quickly move from $u_0$ to $u_f$, and is thus dominated by the conjugate momentum $\mu$, i.e., $\zeta u^2\ll\xi\mu^2$, so that $\mc{H}\sim \xi\mu^2$ as in the parabolic class. For $t\gg\tau_\text{c}$
the cost saturates to
\algn{
	\lim_{t_f\to\infty}\ms{C}^-_{t_f}(u_f, u_0) = \frac{u_f^2 +u_0^2}{2\xi\tau_\text{c}} + \tilde V(u_f)\,,
}
which shows that in the hyperbolic class, finite $u_{f,0}$ and thus finite $\lambda_{f,0}^*$ are penalized even at long times. This is different from the parabolic class discussed previously, but consistent with the expectation for the specific control cost~\eqnref{Cquad} that penalizes any deviation of $\lambda(t)$ from zero, which becomes necessary when $u_{f,0}\neq 0$.

\paragraph*{Elliptic class ($\zeta > 0$).}
The Hamilton equations now describe a harmonic oscillator with period $\tau_\text{c}$. The matrix exponential $\mbb{M}^+_t$ and $\mbb{N}^+_t$ involve trigonometric functions, reading
\algn{
	\mbb{M}^+_t &= \pmat{\cos\left(\frac{t}{\tau_\text{c}}\right)	&	\tau_\text{c}\xi\sin\left(\frac{t}{\tau_\text{c}}\right) \\ -\frac1{\tau_\text{c}\xi}\sin\left(\frac{t}{\tau_\text{c}}\right)	&	\cos\left(\frac{t}{\tau_\text{c}}\right)}\,, \\
	\mbb{N}^+_t &= \frac{1}{\sin\left(\frac{t_f}{\tau_\text{c}}\right)}\pmat{\sin\left(\frac{t_f-t}{\tau_\text{c}}\right)	&	\sin\left(\frac{t}{\tau_\text{c}}\right) \\ -\frac1{\tau_\text{c}\xi}\cos\left(\frac{t_f-t}{\tau_\text{c}}\right)	&	\frac1{\tau_\text{c}\xi}\cos\left(\frac{t}{\tau_\text{c}}\right)}\,.
}
This yields the optimal solutions
\algn{\eqnlab{uell}
    u^*(t) &= \frac1{\sin\left(\frac{t_f}{\tau_\text{c}}\right)}\left[\sin\left(\frac{t}{\tau_\text{c}}\right)u_f + \sin\left(\frac{t_f-t}{\tau_\text{c}}\right)u_0\right]
    \,,\\
     \mu^*(t) &= \frac{1}{\tau_\text{c}\xi\sin\left(\frac{t_f}{\tau_\text{c}}\right)}
    \left[\cos\!\left(\frac{t}{\tau_\text{c}}\right)u_f
    - \cos\!\left(\frac{t_f-t}{\tau_\text{c}}\right) u_0  
    \right]\,,
  }
  and the optimal protocol
  \begin{multline}
	\lambda^*(t) = \frac1{\sin\left(\frac{t_f}{\tau_\text{c}}\right)}\left\{\left[\sin\left(\frac{t}{\tau_\text{c}}\right)+\frac{\tau_\text{p}}{\tau_\text{c}}\cos\!\left(\frac{t}{\tau_\text{c}}\right)\right]u_f\right.\\
	\left. + \left[\sin\left(\frac{t_f-t}{\tau_\text{c}}\right)-\frac{\tau_\text{p}}{\tau_\text{c}}\cos\!\left(\frac{t_f-t}{\tau_\text{c}}\right)\right]u_0\right\}\,,
\end{multline}
for $0<t<t_f$, together with the endpoint values
$\lambda^*(0)=(1-\Delta_0)u_0$ and $\lambda^*(t_f)=(1-\Delta_f)u_f$ of
\Eqnref{lambdabc}. The corresponding jumps of the optimal protocol are
\algn{
	\Delta\lambda_{f,0} = -\frac{\tau_\text{p}}{\tau_\text{c}}\,
	\frac{\cos\left(\frac{t_f}{\tau_\text{c}}\right)u_{f,0} - u_{f,0}}
	{\sin\left(\frac{t_f}{\tau_\text{c}}\right)}
	\mp \Delta_{f,0}\,u_{f,0}\,.
}
which are, again, unequal in magnitude at the two ends, even for $\Delta_{f,0}=0$.
Figures~\figref{protocols}(a)--(c) show $u^*(t)$, $\lambda^*(t)$, and $\mu^*(t)$, respectively, for the elliptic class as the blue lines. Again, the optimal protocol $\lambda^*(t)$  in \Figref{protocols}(b) corresponds to the jump-free case $\Delta\lambda_{f,0}=0$. For the elliptic class, simultaneous translations of $u$ and $\lambda$ are favoured, which explains why both $u^*(t)$ and $\lambda^*(t)$ make larger excursions than in the other classes.

The cost function takes the form
\algn{\eqnlab{Cplus}
\ms{C}^+_{t_f}(u_f, u_0) = \frac1{2\xi\tau_\text{c}}\,
    	\frac{(u_f^2 +u_0^2)\cos\left(\frac{t_f}{\tau_\text{c}}\right)-2u_0u_f}{\sin\left(\frac{t_f}{\tau_\text{c}}\right)}
    	+ \tilde V(u_f)\,.
}
Interestingly, these expressions indicate that the control problem runs into an instability for $t_f\to t_\text{in}=\pi \tau_\text{c}$, indicated by the divergence of the optimal solutions $u^*(t)$, $\lambda^*(t)$, $\mu^*(t)$, and of the cost function $\ms{C}^+_{t_f}$. The times associated with such instabilities are known as \emph{conjugate times} in the mathematical literature on optimal control~\cite{Agr15}. In particular, $\ms{C}^+_{t_f}(u_f, u_0)$
tends to \emph{negative infinity} as $t_f\to t_\text{in}$: For $t_f = t_\text{in}-\delta\tau_\text{c}$ with $\delta\ll1$, we have
$\ms{C}^+_{t_f}(u_f, u_0)\sim (-1)(u_f^2+u_0^2)/(2\xi\tau_\text{c}\delta)\to-\infty$.

The reason for this instability is the positive sign of
$\zeta = -2(C_{11}+2C_{12}+C_{22})>0$, which implies that excursions along the $\lambda^*=u^*$ direction are negatively
penalized, i.e., rewarded. In other words, the system can harvest negative $\ms{C}^+_{t_f}$ by moving to large $u^*$ and $\lambda^*$, ``park'' there to harvest the negative cost, and then return to $(u_f,\lambda_f)$ at the very end. For short times $t<t_\text{in}$, the initial and final excursions outweigh the thus reduced cost. For $t\geq t_\text{in}$, by contrast, moving to infinite $u$ and $\lambda$ becomes optimal and the system is unstable. Consequently, for $t>t_\text{c}$ the seemingly finite-optimal cost in \Eqnref{Cplus} does not correspond to the minimum cost any more but is simply the value at a saddle point that has lost its stability, while the optimal cost is $\ms{C}^+_{t_f}(u_f, u_0)=-\infty$ for $t>t_\text{in}$.
To make this argument more quantitative, consider protocols that drive the system to a displacement $u=\lambda=a$ along the unstable direction, hold it there for a time $t_\text{h}$, and return at the end. Since $\zeta>0$, the running cost accrued while parked is negative and grows as $\sim-\zeta a^2 t_\text{h}$, whereas the cost of reaching $a$ and returning is fixed, $\sim a^2/(\xi t_\text{r})$ with $t_\text{r}$ the short excursion time. The total, $\sim a^2[1/(\xi t_\text{r})-\zeta t_\text{h}]$, is driven to $-\infty$ as $a\to\infty$ once $t_\text{h}$ is large enough for the bracket to turn negative. For the complete system, this becomes possible precisely when $t_f\geq t_\text{in}=\pi\tau_\text{c}$.

This unbounded behavior is a consequence of the linear dynamics and the quadratic cost function. Running into the instability requires $u^*$, $\lambda^*$, and $\mu^*$ to become large, and thus to leave the linearizable regime. We should therefore require not only that $t<t_\text{in}$, but that $t$ remains sufficiently far away from $t_\text{in}$ for the linearized theory to remain applicable. For $t\ll\tau_\text{c}$ the cost again reduces to the parabolic form \Eqnref{C0}, indicating that for short protocols the dynamics is dominated by the conjugate momentum $\mu$, and the form of the $u$-dependent contribution to $\ms{H}$ (determined by $\zeta$) is irrelevant at leading order.

The three triplets of optimal trajectories $[u^*(t)$, $\lambda^*(t)$, $\mu^*(t)]$ for the three equivalence classes and their associated cost functions $\ms{C}^{\eta}_{t_f}(u_f,u_0)$ are the central results of the Hamiltonian formulation.  Each is fully determined by $u_0$, $u_f$, $t_f$, the parameter $\xi$ and the timescale $\tau_\text{c}$ or $\tau_0$. In addition, the magnitudes of the initial and endpoint jumps of $\lambda^*(t)$ depend on $\Delta_{f,0}$.

Figures~\figref{costclasses}(a)--(c) show all cost functions $\ms{C}^\eta_{t_f}(u_f,u_0)$ as functions of $u_f$ for different times for $\tilde V(u_f) = V_\text{DW}(u_f)$ [\Eqnref{dwell-main}]. We observe a crucial aspect of the problem, discussed in detail in the next
section: with increasing $t_f$, the single minimum of $\ms{C}^\eta_{t_f}(u_f,0)$ at $u_f = 0$ splits into two symmetric minima in all three classes. Because we need to minimize $\ms{C}^\eta_{t_f}(u_f,u_0)$ over the free endpoint $u_f$ to obtain the optimal final position $u_f^*$ and the optimal total cost $\ms{C}^{\eta,*}_{t_f}(u_0)$, this splitting is the hallmark of the control transition.
\begin{figure*}
	\includegraphics[width=\linewidth]{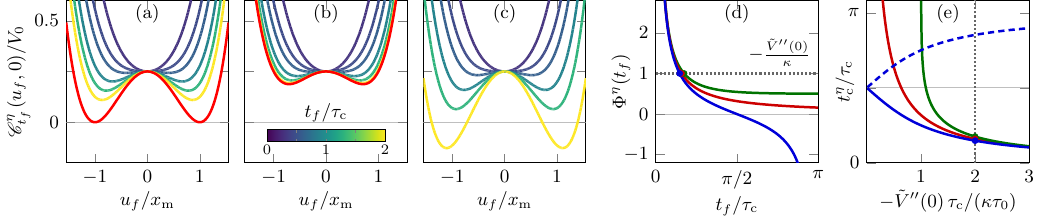}
	\caption{Finite-time control transition in the three
	equivalence classes for $\tilde V = V_\text{DW}$, $\tilde V''(0)=-\kappa$,  and $\tau_\text{c} = 2\tau_0$.
	(a)--(c) Cost function at vanishing initial position for the
	parabolic (a), hyperbolic (b), and elliptic (c) class. The protocol duration
	$t_f$ increases from dark to light [color bar in (b)]. The red lines in (a) and (b) show the cost approached
	as $t_f\to\infty$. The elliptic class in (c) possesses no such limit.
	(d) Cost curvature $\Phi^\eta(t_f)$ of \Eqsref{Phidef} for the
	hyperbolic (green), parabolic (red), and elliptic (blue) class. The transition
	occurs where $\Phi^\eta$ crosses $-\tilde V''(0)/\kappa$ (dotted line),
	marked by dots.
	(e) Critical times~\eqnref{tcs} as a function of the obstacle strength
	$-\tilde V''(0)\tau_\text{c}/(\kappa\tau_0)$. The hyperbolic branch exists only
	beyond unity, where it diverges, in accordance with~\Eqnref{condition2}. For
	the elliptic class the solid line corresponds to $\tilde V''(0)<0$ and the
	dashed line to $\tilde V''(0)>0$ of equal magnitude, for which
	$\pi\tau_\text{c}/2 < t_\text{c}^+ < \pi\tau_\text{c}$. The dotted vertical
	line marks the parameters used in (a)--(d).}
	\figlab{costclasses}
\end{figure*}
%

% =========================================================
\section{Finite-Time Control Transition}
\seclab{transition}
% =========================================================

% -------------------------------------------------------
\subsection{Optimality of the zero-control protocol}
\seclab{stability}
% -------------------------------------------------------

In the final step, we minimize the cost functions $\ms{C}^\eta_{t_f}(u_f, u_0)$ derived in \Secref{hamclass} over the free final position $u_f$ to obtain the optimal total cost
\algn{\eqnlab{optcost}
    \ms{C}^{\eta,*}_{t_f}(u_0) = \min_{u_f}\,\ms{C}^\eta_{t_f}(u_f, u_0)\,,
}
and the corresponding optimal final position $u_f^* = u_f^*(u_0, t)$ that achieves it.
The qualitative character of this minimization depends critically on whether
$\ms{C}^\eta_{t_f}(u_f, u_0)$, viewed as a function of $u_f$ at fixed $u_0$ and
$t_f$, has a unique minimum or multiple local minima, see \Figref{costclasses}(a)--(c). We show that this
character changes at a critical time $t_\text{c}^\eta$, producing a transition in
the optimal control strategy.

\paragraph*{Optimality of the zero-control protocol.}
To identify when the zero-control protocol $\lambda^*(t) = 0$ (equivalently,
$u_f^* = 0$ for $u_0 = 0$) is optimal, we expand
$\ms{C}^\eta_{t_f}(u_f, 0)$ around $u_f = 0$ for $u_0 = 0$. We assume here that
$\tilde V(u_f)$ is even, $\tilde V(u_f) = \tilde V(-u_f)$, i.e., the obstacle is symmetric about the zero-control position.
This significantly simplifies the analysis, but need not be the case in general. However, as we show in
Appendix~\secref{Aasym}, a \emph{symmetry-breaking} $\tilde V$ leaves the picture qualitatively
unchanged.

For even $\tilde V(u_f)$, the second-order expansion contains only even powers of $u_f$:
\algn{\eqnlab{expand}
    \ms{C}^\eta_{t_f}(u_f, 0) \sim \tilde V(0)
    + \left[\kappa\,\Phi^\eta(t) + \tilde V''(0)\right]\frac{u_f^2}2\,,
}
where $\tilde V''(0) = \ddd{u_f}\tilde V(u_f)\big|_{u_f=0}$ and we defined the time-dependent cost curvature
\sbeqs{\eqnlab{Phidef}
\algn{
    \Phi^0(t_f) &= \frac{\tau_0}{t_f}\,,\\
    \Phi^-(t_f) &= \frac{\tau_0}{\tau_\text{c}}\coth\left(\frac{t_f}{\tau_\text{c}}\right)\,,\\
    \Phi^+(t_f) &= \frac{\tau_0}{\tau_\text{c}}\cot\left(\frac{t_f}{\tau_\text{c}}\right)\,,
}
}
for the parabolic, hyperbolic, and elliptic classes, respectively.

Figure~\figref{costclasses}(d) shows $\Phi^\eta(t_f)$ for the three classes. As discussed previously, the curvatures $\Phi^\pm(t_f)$ in the elliptic and hyperbolic classes approach $\Phi^0(t_f)$ for short times ($t\ll\tau_{0,\text{c}}$). 
In this limit, all $\Phi^\eta(t_f)$ become large, reflecting the high, momentum-dominated cost of any displacement from $u_f = 0$
when the protocol is short. As $t_f$ increases, all $\Phi^\eta(t_f)$ decrease monotonically in the following ways:
\begin{itemize}
    \item For the parabolic class, $\Phi^0(t_f) = \tau_0/t_f$ decays
    to zero algebraically as $t_f\to\infty$.
    \item For the hyperbolic class, $\Phi^-(t_f)$ decays exponentially from
    $\tau_0/t_f$ at short times to the constant value $\tau_0/\tau_\text{c}$
    at long times.
    \item For the elliptic class, $\Phi^+(t)$ decays from $\tau_0/t_f$ to
    zero at $t_f = \pi\tau_\text{c}/2$ and diverges to negative infinity at $t=\pi\tau_\text{c}$.
    For times $t_f>\pi\tau_\text{c}$ the linear control problem for the elliptic class is unstable,
    as explained in the previous section.
\end{itemize}
The zero-control protocol corresponds to $u_f^* = 0$, which is a
stationary point of $\ms{C}^\eta_{t_f}(u_f, 0)$ for all $t_f$ by the symmetry of
$\tilde V$. Its stability is governed by the sign of the coefficient of
$u_f^2$ in \Eqnref{expand}. When this coefficient is positive, $u_f = 0$
is a local minimum and the zero-control protocol is (locally) optimal.
When the coefficient is negative, $u_f = 0$ becomes a local maximum and
a non-trivial optimal final position $u_f^* \neq 0$ emerges.

\paragraph*{Necessary conditions for the transition.}
The coefficient of $u_f^2$ in \Eqnref{expand} changes sign when
$\kappa\Phi^\eta(t_\text{c}^\eta) + \tilde V''(0) = 0$, i.e., when 
\algn{\eqnlab{tcdef}
	\Phi^\eta(t_\text{c}^\eta) = -\frac{\tilde V''(0)}{\kappa}\,.
}
For this to occur at a finite positive time $t_\text{c}^\eta$, there are different possibilities, depending on $\eta$.
For the parabolic class $\eta=0$, we have $\Phi^0(t_f)>0$, so the right-hand side of \Eqnref{tcdef} needs to be positive, which in turn requires
\algn{\eqnlab{condition}
    \tilde V''(0) < 0\,.
}
The condition \eqnref{condition} implies that the noise-averaged penalty $\tilde V$ must have a local
\emph{maximum} at the zero-control fixed point $u_f = 0$, i.e., the obstacle
must be hardest to cross at the point toward which the zero-control
protocol steers the particle~\cite{Mei26b}.

For the hyperbolic $\eta=-1$ class, $\Phi^-(t_f)$ monotonically decreases to the value $\tau_0/\tau_\text{c}$, so we must require
\algn{\eqnlab{condition2}
	\tilde V''(0)<-\frac{\kappa \tau_0}{\tau_\text{c}}=-\left|\frac{\zeta}{\xi}\right|^{1/2}\,.
}
Hence for $\eta=-1$, the stronger condition~\eqnref{condition2} demands that
the curvature of $\tilde V$ at $u_f=0$ be smaller than a negative threshold that depends on $\zeta$ and $\xi$.

For the elliptic class, $\Phi^+(t_f)$ monotonically decreases from positive infinity to negative infinity, so \Eqnref{tcdef} can be inverted for all $\tilde V''(0)$. The sign of $\tilde V''(0)$ determines whether $t^+_\text{c}<\pi\tau_\text{c}/2$ for $\tilde V''(0)<0$ or $\pi\tau_\text{c}/2<t_\text{c}^+<\pi\tau_\text{c}$ for $\tilde V''(0)>0$.

If the conditions \eqnref{condition} or \eqnref{condition2} for the respective classes $\eta=0$ or $\eta=-1$ are not fulfilled, the coefficient of $u_f^2$ in \Eqnref{expand} remains positive for all $t > 0$, which implies that the zero-control protocol remains optimal for all protocol durations, and no finite-time control transition occurs. In the elliptic class, finite-time control transitions always occur, independently of the details of $\tilde V$.

\paragraph*{Critical times.}
Based on these considerations, the critical times $t_\text{c}^\eta$ at which the
zero-control protocol becomes unstable are given by
\sbeqs{\eqnlab{tcs}
\algn{
    \frac{t_\text{c}^0}{\tau_0} &= -\frac{\kappa}{\tilde V''(0)}\,,
    \eqnlab{tc0}\\[6pt]
    \frac{t_\text{c}^-}{\tau_\text{c}} &= \arccoth\!\left(-\frac{\tilde V''(0)}{\kappa}\frac{\tau_\text{c}}{\tau_0}\right)\,,
    \eqnlab{tcminus}\\[6pt]
    \frac{t_\text{c}^+}{\tau_\text{c}} &= \arccot\!\left(-\frac{\tilde V''(0)}{\kappa}\frac{\tau_\text{c}}{\tau_0}\right)\,.
    \eqnlab{tcplus}
}
}
For a given $\tilde V$ and under the conditions \eqnref{condition} and \eqnref{condition2} for $\eta =0$ and $\eta=-1$, respectively, there is a unique, single time at which the zero-control protocol loses stability. For the elliptic class, by contrast, no additional condition is necessary. Instead a unique critical time $t_\text{c}^+$ exists for any sign of $\tilde V$, but we have $0<t_\text{c}^+<\pi\tau_\text{c}/2$ for $\tilde V''(0)<0$ and $\pi\tau_\text{c}/2<t_\text{c}^+<\pi\tau_\text{c}$ for $\tilde V''(0)>0$ as discussed above.
Figure~\figref{costclasses}(e) shows the critical times
\Eqnref{tcs} as a function of the obstacle strength, including both elliptic
branches (blue).

\paragraph*{Connection to dynamical timescales.}
The critical times \Eqnref{tcs} are set by the competition between the \textit{cost curvature} $\kappa\Phi^\eta$ and the \textit{penalty curvature}
$\tilde V''(0)$ and depend on both control timescales $\tau_0$ and $\tau_\text{c}$, as well as the obstacle strength $|\tilde V''(0)|$ relative to $\kappa$.
For weak obstacles [i.e., small $|\tilde V''(0)|\tau_\text{0}/(\kappa\tau_{c})$], the critical times are
larger than the dynamical timescales $\tau_\text{c}$ (or $\tau_0$), so that the zero-control
protocol is optimal over a wide range of durations $t_f$. For stronger obstacles ($|\tilde V''(0)|\sim\kappa\tau_\text{c}/\tau_{0}$), the critical time is of the order of $\tau_\text{c}$ or $\tau_0$, and the transitions occur within the regime where the full control dynamics are relevant. The
experiments described in Ref.~\cite{Mei26b} and \Secref{experiments} operate in this latter
regime.

\paragraph*{Behaviour for non-zero initial condition, $u_0\neq 0$.}
The analysis above is restricted to $u_0 = 0$, where the symmetry of
$\tilde V$ ensures that $u_f = 0$ is a stationary point of
$\ms{C}^\eta_{t_f}(u_f, 0)$. For $u_0\neq 0$ the symmetry is broken and \Eqnref{expand} picks up uneven
terms in $u_f$. Consequently, $u_f = 0$ is no longer a minimizer of $\ms{C}^\eta(u_f,u_0)$.

In the next section, we show that the minimizer $u_f^*(u_0, t)$ is a smooth
function of $u_0$ for $t _f< t_\text{c}^\eta$, but transitions discontinuously as
$u_0$ passes through zero for $t_f > t_\text{c}^\eta$. This discontinuous transition
of $u_f^*(u_0, t_f)$ at $u_0 = 0$ produces a kink in the optimal cost
$\ms{C}^{\eta,*}_{t_f}(u_0)$ at $u_0 = 0$. This kink in turn is an observable signature of the control transition and
the quantity measured in our optimal control experiment.

% -------------------------------------------------------
\subsection{Topology of the cost landscape and the Landau analogy}
\seclab{landau}
% -------------------------------------------------------

The stability analysis of \Secref{stability} identifies when the
zero-control protocol becomes suboptimal, but does not describe the full
structure of the cost landscape $\ms{C}^\eta_{t_f}(u_f, u_0)$ as a function of
$u_f$ for all times $t_f$ and initial positions $u_0$. We now examine this
structure in more detail, drawing on the analogy with the Landau theory of
continuous phase transitions~\cite{Cha95}. Throughout this subsection we focus on the
parabolic class $\ms{C}^0_{t_f}$ for concreteness. The hyperbolic and elliptic
classes differ only in the precise form of $\Phi^\eta(t_f)$ in \Eqsref{Phidef} and the
qualitative picture is the same.

\paragraph*{Topology of the cost function at $u_0 = 0$.}
For $u_0 = 0$ and a symmetric penalty that obeys the condition $\tilde V''(0) < 0$ [\Eqnref{condition}],
the cost $\ms{C}^0_{t_f}(u_f, 0)$ is an even function of $u_f$. Upon Taylor expanding up to fourth order in $u_f$, it takes the
form
\algn{\eqnlab{Cexpand}
    \ms{C}^0_{t_f}(u_f, 0) \sim \tilde V(0)
    + \left[\frac{\kappa\tau_0}{t_f} + \tilde V''(0)\right]\frac{u_f^2}2
    + \tilde V^{(4)}(0)\frac{u_f^4}{4!} \,,
}
where $\tilde V^{(4)}(0)\equiv\ed^4\tilde V/\ed u_f^4|_{u_f=0} > 0$ for a
double-well penalty with minima away from the origin, ensuring the cost is
bounded below. The coefficient of $u_f^2$ changes sign at $t_f = t_\text{c}^0$
given by \Eqnref{tc0}, while the quartic coefficient remains positive
throughout. As already shown previously in \Figref{costclasses}(a), the topology of $\ms{C}^0_{t_f}(u_f, 0)$ as a function of $u_f$
therefore undergoes the following sequence:
\begin{itemize}
    \item For $t_f < t_\text{c}^0$: the quadratic coefficient is positive and the
    cost has a unique minimum at $u_f = 0$. The zero-control protocol is
    optimal.
    \item At $t_f = t_\text{c}^0$: the quadratic coefficient vanishes. The cost
    has a flat direction at $u_f = 0$, controlled entirely by the quartic
    term. The cost is still minimal at $u_f = 0$, but the curvature
    vanishes. The system is at the critical point of the transition.
    \item For $t_f > t_\text{c}^0$: the quadratic coefficient is negative and
    $u_f = 0$ becomes a local maximum. For even $\tilde V(u_f)$, two symmetry-related minima at
    $u_f = \pm u_f^*(t_f)$ emerge continuously from zero. The optimal
    strategy is to steer the particle toward one of these minima,
    avoiding the obstacle at $u_f = 0$.
\end{itemize}
This topology change is directly analogous to the transition of the
Landau free energy from a single-well to a double-well form at a continuous
phase transition. The cost $\ms{C}^0_{t_f}(u_f, 0)$ plays the role of the
Landau free energy $F(m)$, the final position $u_f$ plays the role of the
order parameter $m$, and the protocol duration $t_f$ plays the role of the
inverse temperature $\beta=(\kb T)^{-1}$. The resulting landscape is shown again in \Figref{costfunc}(a).

\begin{figure}[t]
    \centering
    \includegraphics[width=\linewidth]{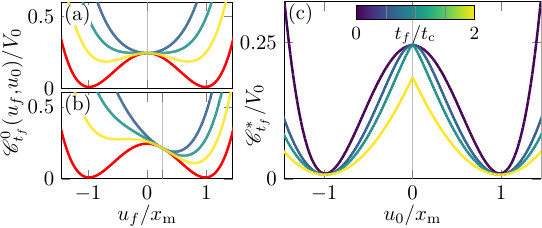}
    \caption{Cost landscape of the finite-time control transition in the parabolic
    class. (a) $\ms{C}^0_{t_f}(u_f, 0)$ as function of $u_f$ for increasing
    protocol duration $t_f$ [color bar in (c)] and the bare obstacle $V_\text{DW}$ (red).
    (b) Same as (a) but for a non-zero initial position $u_0 = 0.25\,x_\text{m}$, breaking the
    $u_f\to-u_f$ symmetry. (c) Minimized cost $\ms{C}^{0,*}_{t_f}(u_0)$ as a function of
    $u_0$, developing a kink at $u_0 = 0$ for $t_f > t_\text{c}^0$.}
    \figlab{costfunc}
\end{figure}

\paragraph*{Order parameter and mean-field exponent.}
To deepen the analogy with equilibrium phase transitions, we define an order parameter of the control transition as the magnitude
of the optimal final position at infinitesimally positive $u_0$,
\algn{\eqnlab{orderpar}
    u_f^* \equiv \lim_{u_0\to 0^+} u_f^*(u_0, t_f)\,.
}
For $t_f < t_\text{c}^0$ we have $u_f^* = 0$, while for $t_f > t_\text{c}^0$ the order
parameter is finite, with a spontaneously broken sign. To determine how $u_f^*$ grows above $t_\text{c}^0$, we
minimize \Eqnref{Cexpand} over $u_f$. Setting the derivative to zero we obtain as the non-trivial solution ($u^*_f \neq 0$) for $t>t^0_\text{c}$
\algn{\eqnlab{utstar}
    (u_f^*)^2 &= -\frac{3!}{\tilde V^{(4)}(0)}
    \left[\frac{\kappa\tau_0}{t_f} + \tilde V''(0)\right]
    \\
   & \sim \frac{3!{\tilde V''(0)}^2}{\kappa\tilde V^{(4)}(0)}\cdot
    \left(\frac{t_f - t_\text{c}^0}{\tau_0}\right)\,,
}
for $|t_f - t_\text{c}^0|\ll \tau_0$, so that close to the critical time
\algn{\eqnlab{mfexp}
    u_f^* \propto \left(t_f - t_\text{c}^0\right)^{1/2}\,,\qquad t_f\to (t_\text{c}^0)^+\,.
}
For the double-well penalty~\eqnref{dwell-main}, the Landau coefficients read explicitly [with the dimensionless noise parameter $\eps$ defined in \Eqnref{epsdef}]
\algn{\eqnlab{Vtildederivs}
    \tilde V_\text{DW}''(0) &= -\frac{V_0}{x_\text{m}^2}(1 - 3\eps)\,,\nn\\
    \tilde V_\text{DW}^{(4)}(0) &= \frac{3!V_0}{x_\text{m}^4}\,,
}
so that the order parameter just above the critical time reads
\algn{\eqnlab{utstarexplicit}
    \left(\frac{u_f^*}{x_\text{m}}\right)^2 \sim \Gamma_\text{DW}(1 - 3\eps)^2\left(\frac{t_f - t_\text{c}^0}{\tau_0}\right)\,,
    \qquad t_f\to(t_\text{c}^0)^+\,,
}
with the dimensionless parameter 
\algn{\eqnlab{Gamdef}
	\Gamma_\text{DW} = \frac{V_0}{\kappa x_\text{m}^2}\,,
}
which determines the magnitude of the obstacle relative to that of the harmonic trap.
The order parameter $u_f^*$ grows as the square root of the distance from the
critical time, leading to the mean-field critical exponent $\beta_\text{MF} =
1/2$. This result holds for all three classes and follows entirely from
the symmetry of $\tilde V$ and the smoothness of $\ms{C}^\eta_{t_f}$, without
reference to the specific form of the cost. It is the control analogue of
the mean-field magnetization $m\propto(\beta - \beta_c)^{1/2}$ in the
Curie-Weiss model~\cite{Mei22a}.

\paragraph*{Effect of a non-zero initial position.}
When $u_0\neq 0$ the symmetry $u_f\to-u_f$ of the cost is broken, and the
two potential minima are no longer degenerate, as shown in \Figref{costfunc}(b). The situation is directly
analogous to applying an external field $H\neq 0$ to a magnetic
system close to an equilibrium phase transition: one minimum is deepened and the other raised, so the optimizer
selects a unique $u_f^*$ for every $u_0\neq 0$. For $t_f > t_\text{c}^0$, the optimal final position $u_f^*(u_0, t_f)$ varies continuously
with $u_0$ except at $u_0 = 0$, where it jumps discontinuously between the
two branches $\pm|u_f^*|$. This is the analogue of the discontinuous jump
of the equilibrium magnetization at $H = 0$ below the critical temperature.

\paragraph*{Kink in the optimal cost.}
The discontinuous jump of $u_f^*(u_0, t_f)$ at $u_0 = 0$ for $t_f > t_\text{c}^\eta$
has a direct consequence for the optimal cost $\ms{C}^{\eta,*}_{t_f}(u_0)$
defined in \Eqnref{optcost}. Since $u_f^*$ transitions between two
branches at $u_0 = 0$, the optimal cost develops a \emph{kink} at $u_0 = 0$
for $t_f > t_\text{c}^\eta$, shown in Fig.~\ref{fig:costfunc}(c). At the kink, the left and right derivatives of $\ms{C}^{\eta,*}_{t_f}$ with
respect to $u_0$ are equal in magnitude but opposite in sign. For $t_f < t_\text{c}^\eta$,
by contrast, $u_f^*(u_0,t_f)$ is a smooth function of $u_0$ and
$\ms{C}^{\eta,*}_{t_f}(u_0)$ is smooth. The transition from smooth to kinked
optimal cost at $t_f = t_\text{c}^\eta$ is a directly observable signature of
the control transition, since $\ms{C}^{\eta,*}_{t_f}(u_0)$ is the quantity
measured in the experiments described in \Secref{experiments}.

In the Landau analogy, the kink in $\ms{C}^{\eta,*}_{t_f}(u_0)$ corresponds to
the kink in the equilibrium free energy $F(H)$ at $H = 0$ below $T_c$.
Both arise from a Maxwell construction~\cite{Cha95} selecting the lower of two
degenerate branches. The correspondence between the control problem and
the equilibrium phase transition is summarized in Table~\ref{tab:landau}.

In the present analysis we have assumed for simplicity that $\tilde V(u_f)$ is even, $\tilde V(u_f) = \tilde V(-u_f)$,
i.e., the obstacle is symmetric about the zero-control position. We detail in Appendix~\secref{Aasym}
how this picture changes quantitatively when $\tilde V$ is not even. In either case the optimal final position $u^*_f$ jumps across a first-order line in the
$u_0$-$t_f$ plane, which terminates at a critical point where the transition is
continuous. The $u_f\to-u_f$ symmetry merely pins the first-order line (and the critical point) to $u_0 = 0$.
In general, the line is a non-trivial curve in the $(u_0,t_f)$ plane with $u^*_0\neq0$ at the critical point,
while the qualitative picture remains the same.

\begin{table*}[t]
\caption{Correspondence between the finite-time control transition and the
continuous phase transition of an equilibrium spin model (e.g., the
Curie-Weiss model~\cite{Mei22a}).}
\label{tab:landau}
\begin{ruledtabular}
\begin{tabular}{lll}
\textbf{Concept} & \textbf{Control problem} & \textbf{Equilibrium transition} \\
\hline
Control parameter & Protocol duration $t_f$ & Inverse temperature $\beta$ \\
Critical value     & Critical time $t_\text{c}^\eta$ & Critical temperature $\beta_c$ \\
Order parameter    & Optimal final position $u_f^*$ & Magnetization $\bar m$ \\
External field     & Initial position $u_0$ & External field $H$ \\
Free energy        & Optimal cost $\ms{C}^{\eta,*}_{t_f}(u_0)$ & Free energy $F(H)$ \\
Landau potential   & Cost $\ms{C}^\eta_{t_f}(u_f, u_0)$ & Constrained free energy $F(m,H)$ \\
Critical exponent  & $u_f^*\propto(t_f-t_\text{c}^\eta)^{1/2}$ & $\bar m\propto(\beta-\beta_c)^{1/2}$ \\
Symmetry           & $u_f\to -u_f$ & $m\to -m$ (Ising) \\
\end{tabular}
\end{ruledtabular}
\end{table*}

The ``mean-field'' description of the control transition is exact for the
optimal control problem, because the cost $\ms{C}^\eta_{t_f}(u_f, u_0)$ is a
deterministic function of $u_f$ and no fluctuations of the order parameter
$u_f^*$ occur. This is in contrast to equilibrium phase transitions, where the order parameter,
even at mean-field level, shows strong fluctuations close to the transition, characterized by critical exponents.
The absence of fluctuations in the control transition means that
the sharp transition shown in \Figref{costfunc}(c) is not a (weak-noise) approximation,
but the exact description of the system at any noise level.
Consequently, the kink in $\ms{C}^{\eta,*}_{t_f}(u_0)$ is perfectly sharp, with no rounding, while non-analytic points in the free energy at equilibrium phase transitions can be sharp only in the thermodynamic limit~\cite{Cha95}. 

% -------------------------------------------------------
Finally, \Eqsref{lambdastar} and \eqnref{Deltalam} show how the finite-time control transition affects
the optimal protocol $\lambda^*(t)$. For $u_0 = 0$ and $t_f < t_\text{c}^0$ the optimal final position
vanishes, so $u^*(t)=\mu^*(t)=0$. Consequently, also the optimal protocol $\lambda^*(t)$ is identically zero, i.e., no steering
occurs. For $t_f > t_\text{c}^0$ the final position bifurcates to
$u_f^* = \pm|u_f^*|$ and $\lambda^*(t)$ becomes non-trivial with boundary
jumps of magnitude $|u_f^*\tau_\text{p}/t_f+\Delta_{f,0}u_{f,0}|$, for all equivalence classes of \Secref{hamclass}. The onset of a non-zero optimal protocol amplitude at
$t_\text{c}^\eta$ is thus the protocol-level signature of the control transition.

% =========================================================
\section{Connection to Finite-Time Dynamical Phase Transitions}
\seclab{mapping}
% =========================================================

% -------------------------------------------------------
\subsection{Mapping between optimal control and non-equilibrium relaxation}
\seclab{mapdetail}
% -------------------------------------------------------

The cost functions $\ms{C}^\eta_{t_f}(u_f, u_0)$ derived in \Secref{hamclass}
bear a striking structural resemblance to the large-deviation rate
functions that govern rare fluctuations in non-equilibrium relaxation
processes. In this section we make this resemblance precise by
establishing a mapping between the two problems,
identify the physical origin of the correspondence, and determine for
which equivalence classes the mapping holds.

\paragraph*{Large deviations in non-equilibrium relaxation.}
As a prototypical relaxation problem, consider a Brownian particle initially at thermal equilibrium in a
confining potential $V_\text{eq}(x)$. At time $t = 0$ the potential is
either replaced by a new potential $V_\text{q}(x)$, or removed entirely.
Such a protocol is known as a \emph{potential quench}. The particle then
relaxes into the new potential, or diffuses freely for a time $t_f$, starting from
the equilibrium distribution $P_\text{eq}(x_0)\propto
\ee^{-V_\text{eq}(x_0)/(\kb T)}$ and ending at some position $x_f = x(t_f)$.

For weak noise $\kb T\ll V_0$, with $V_0$ a characteristic
energy scale of $V_\text{eq}$ [for the double well~\eqnref{dwell-main}, $V_0$ is
proportional to the barrier height], the probability of
finding the particle at position $x_f$ at time $t_f$ takes the large-deviation
form~\cite{Tou09,Fre84}
\algn{\eqnlab{LDform}
    P(x_f, t_f) \propto \exp\!\left[-\frac{\ms{V}^*_{t_f}(x_f)}{\kb T}\right]\,,
}
where $\ms{V}^*_{t_f}(x_f)$ denotes the \emph{rate function}. This rate function is obtained by first minimizing
the action of the path probability over all intervening trajectories $x(t)$ with fixed endpoints $x(0)=x_0$
and $x(t_f)=x_f$, and then minimizing over the initial position $x_0$:
\algn{\eqnlab{ratefunc}
    \ms{V}^*_{t_f}(x_f) &= \min_{x_0}\,\ms{V}_{t_f}(x_f, x_0)\,,
    \\
    \ms{V}_{t_f}(x_f, x_0) &= \!\!\!\!\!\!\!\!\!\!\!\!\!\!\!\!\!\min_{\substack{x(t):\\x(0)=x_0,\,x(t_f)=x_f}}\!\!\!\!\!\!
    \left\{\int_0^{t_f}\!\!\ed t\,\ms{L}[x(t),\dot x(t)] + V_\text{eq}(x_0)\right\}.
}
The functional inside the braces is the so-called Onsager--Machlup action~\cite{Ons53},
and $\ms{L}[x(t),\dot x(t)]$ is the Lagrangian for the post-quench dynamics.
For a quench into a harmonic potential
$V_\text{q}(x) = {\kappa_\text{q}}x^2 /2$ with friction coefficient $\gamma$,
where the subscript distinguishes the post-quench stiffness
$\kappa_\text{q}$ from the stiffness $\kappa$ of the control trap in
\Eqnref{meandyn}, the Lagrangian takes the quadratic form
\algn{\eqnlab{Lagrangian}
    \ms{L}[x(t),\dot x(t)] = \frac{\gamma}{4}\!\left[\dot x(t)
    + \frac{\kappa_\text{q}}{\gamma}x(t)\right]^2\,,
}
and for a quench into free diffusion ($\kappa_\text{q} = 0$) it reduces to
$\ms{L} = \gamma\dot x^2/4$.

\paragraph*{Hamiltonian structure of the relaxation problem.}
The minimization over trajectories in \Eqnref{ratefunc} is a variational
problem that can be cast in Hamiltonian form by the Legendre transform of
$\ms{L}$. Introducing the conjugate momentum $p(t) = \partial\ms{L}/\partial\dot x$,
the Hamiltonian for the post-quench dynamics reads
\algn{\eqnlab{HR}
    \ms{H}_R(x,p) = p\dot x - \ms{L}[x,\dot x]
    = \frac1{\gamma}\left(p^2 - \kappa_\text{q}xp\right)\,,
}
and the rate function $\ms{V}_{t_f}(x_f, x_0)$ for fixed endpoints is
\algn{\eqnlab{Vfixed}
    \ms{V}_{t_f}(x_f, x_0) = \int_0^{t_f}\!\!\ed t\,\Big[p(t)\dot x(t)
    - \ms{H}_R(x,p)\Big] + V_\text{eq}(x_0)\,,
}
where $x(t)$ and $p(t)$ satisfy the Hamilton equations
\algn{\eqnlab{HamR}
    \dot x(t) = \deldel{p}\ms{H}_R = \frac1{\gamma}\left(2p
    - \kappa_\text{q}x\right)\,,
    \quad
    \dot p(t) = -\deldel{x}\ms{H}_R = \frac{\kappa_\text{q}}{\gamma}p\,,
}
with boundary conditions $x(0) = x_0$ and $x(t_f) = x_f$. The
Hamiltonian matrix $\Hmat_R$ associated with the quadratic form
$\ms{H}_R = \frac{1}{2}\boldsymbol{q}^{\sf T}\Hmat_R\boldsymbol{q}$ [where
$\boldsymbol{q} = (x,p)^{\sf T}$] is
\algn{\eqnlab{HmatR}
    \Hmat_R = \frac{1}{\gamma}\pmat{0 & -\kappa_\text{q} \\ -\kappa_\text{q} & 2}\,.
}
In its current form, $\Hmat_R$ is not diagonal but it is brought to diagonal form by
the same canonical transformation \Eqnref{Smat} that implements the gauge
freedom of the control problem. Demanding that the off-diagonal element of
$\Hmat'_R=\Smat^{\sf T}\Hmat_R\Smat$ vanish gives
\algn{\eqnlab{alphadiag}
	\alpha^*_R = \frac{\kappa_\text{q}}{2}\,,
}
and thus
\sbeqs{
\algn{\eqnlab{HmatRdiag}
    \Hmat'_R = \pmat{\zeta_R & 0 \\ 0 & \xi_R}\,,
}
with
\algn{\eqnlab{zetxiR}
	\zeta_R = -\frac{\kappa_\text{q}^2}{2\gamma}\,,\quad
	\xi_R = \frac{2}{\gamma}\,.
}
}
Consequently, by \Eqnref{tauc} and the definition of $\tau_0$ below
\Eqnref{H0normal}, the relaxation problem carries the timescales
\algn{\eqnlab{tauR}
	\tau_\text{c}^R = |\zeta_R\xi_R|^{-1/2} = \tau_\text{p}^R
	\equiv\frac{\gamma}{\kappa_\text{q}}\,,\qquad
	\tau_0^R = \frac{1}{\kappa_\text{q}\xi_R} = \frac{\tau_\text{p}^R}{2}\,.
}
The relaxation problem therefore always sits at the fixed timescale ratio~\footnote{Note that the post-quench relaxation time
$\tau_\text{p}^R = \gamma/\kappa_\text{q}$ of \Eqnref{tauR} is a property of
the relaxation problem and differs in general from the trap relaxation time
$\tau_\text{p} = \gamma/\kappa$ of the control problem. The two coincide only
when $\kappa_\text{q} = \kappa$, i.e.\ for $\tau_0/\tau_\text{c} = 1/2$ in
\Eqnref{kappaq}.} $\tau_0^R/\tau_\text{c}^R = 1/2$, whereas in the control problem this ratio is
free.

An equivalent way to diagonalize $\mbb{H}_R$ is by expanding the square in
\Eqnref{Lagrangian} and isolating the total derivative:
\algn{\eqnlab{OMLexpand}
	\ms{L}[x,\dot x] = \frac{\gamma}{4}\left[\dot x^2
	+ \left(\frac{x}{\tau_\text{p}^R}\right)^2\right]
	+ \frac{\kappa_\text{q}}{4}\dd{t}x^2\,,
}
so that
\algn{\eqnlab{OMsplit}
	\int_0^{t_f}\!\!\ed t\,\ms{L}[x,\dot x] = \int_0^{t_f}\!\!\ed t\,\Big[p(t)\dot x(t)
    - \ms{H}'_R(x,p)\Big]
	+ \frac{\kappa_\text{q}}{4}\left(x_f^2 - x_0^2\right)\,,
}
where $\ms{H}'_R(x,p) = (\xi_R p^2 + \zeta_R x^2)/2$, consistent with \Eqnref{HmatRdiag}. The determinant of $\Hmat_R$ (and $\Hmat'_R$) is
\algn{\eqnlab{detHR}
    \det\Hmat_R = -\frac{\kappa_\text{q}^2}{\gamma^2} \leq 0\,,
}
which is negative for $\kappa_\text{q} > 0$ (hyperbolic class) and zero for
$\kappa_\text{q} = 0$ (parabolic class). The Hamiltonian $\Hmat_R$ therefore
covers only the parabolic and hyperbolic classes but not the elliptic
class. The elliptic class
($\det\Hmat > 0$) would require $\det\Hmat_R > 0$, which from
\Eqnref{detHR} would necessitate $\kappa_\text{q}^2 < 0$, i.e.\ an imaginary
trap stiffness. We therefore conclude that the elliptic control class has 
no simple relaxation counterpart.

\paragraph*{The mapping.}
Comparing \Eqnref{Vfixed} with the control cost
\Eqnref{Hform}--\Eqnref{Copt}, the two expressions have the same
mathematical form: both are functionals of a quadratic Hamiltonian
with fixed endpoints, plus a boundary penalty. The duality between the two
problems is established by the following identifications:
\begin{itemize}
    \item The mean position $u(t)$ maps to the particle position $x(t)$,
    and the conjugate momentum $\mu(t)$ maps to $p(t)$.
    \item The noise-averaged final penalty $\tilde V(u_f)$ maps to the initial potential plus a harmonic shift
    $V_\text{eq}(x_0)-\kappa_\text{q}x_0^2/4$, i.e., the penalty on the \emph{final} state of the
    control problem becomes a penalty on the \emph{initial} state of the
    relaxation problem.
    \item The optimization over the free final position $u_f$ in the
    control problem maps to the optimization over the initial position
    $x_0$ in the relaxation problem.
    \item Time is reversed \emph{and rescaled} using
    $t\to s\,(t_f-t)$ with the factor $s=\xi_R/\xi$ and $x(t)\to u(t_f-t)$, $p(t) \to -\mu(t_f-t)$.
\end{itemize}
The time rescaling by $s$ is required to match the Hamiltonian matrices
$\Hmat$ and $\Hmat_R$, because it rescales $\Hmat\to s\Hmat$.
In particular, $s$ is fixed by matching the lower-right diagonal
entries, $s\xi=\xi_R$. The upper-left entries then match only under the
condition
\algn{\eqnlab{equivcond}
	\frac{\zeta_R}{\xi_R} = \frac{\zeta}{\xi}\,.
}
Since $\text{sgn}(\zeta/\xi) = \eta$, \Eqnref{equivcond} trivially requires that
the control and relaxation problems belong to the same class. This condition is fulfilled automatically for $\zeta=\zeta_R=0$ (the parabolic class, see also \cite{Mei26b}).
However, for the hyperbolic class, \Eqnref{equivcond} restricts the control problems that are compatible with
a given relaxation problem and vice versa. For $\zeta,\zeta_R\neq 0$, we use \eqnref{equivcond} to rewrite $s$ as
\algn{\eqnlab{sfactor}
	s = \frac{\xi_R}{\xi} = \frac{\zeta_R}{\zeta}
	= \frac{\tau_\text{c}}{\tau_\text{p}^R}\,,
}
which stipulates that for transformations from the relaxation to the control problem within this class,
the control timescale $\tau_\text{c}$ takes the role of the particle relaxation time $\tau_\text{p}^R$.
In principle, additional rescalings of the form~\eqnref{Srescale} could be used to remove the condition~\eqnref{equivcond}
up to a sign that fixes the equivalence class. However, such rescalings would deform the
obstacle [see the remark below \Eqnref{H0normal}], which is why we disregard them here.

Substituting the mapping into \Eqnref{Vfixed} and comparing with
\Eqsref{Hform} and \eqnref{Gbound}, we find that the relaxation problem maps onto a control problem with terminal cost $V_\text{eq}$ and
\algn{\eqnlab{BR}
	\ms{B}_f = \ms{B}_0 = -\frac{\kappa_\text{q}}{4}\,,
}
which corresponds to the boundary term of \Eqnref{OMsplit} split between the two
ends. To reduce it to the standard form \Eqnref{Copt} with
$\ms{B}_{f,0} = 0$, we proceed as in \Secref{pontryagin} by absorbing $\ms{B}_f$ into the terminal cost and
carrying along $\ms{B}_0$ additively, which gives
\sbeqs{\eqnlab{trafos}
\algn{
    \tilde V(x_0) &= V_\text{eq}(x_0)-\frac{\kappa_\text{q}}{4}x_0^2\,,
    \eqnlab{VtotV}\\
    \ms{V}^\eta_{t_f}(x_f, x_0) &= \ms{C}^\eta_{s\,t_f}(x_0, x_f)
    +\frac{\kappa_\text{q}}{4}x_f^2\,.
    \eqnlab{VtoC}
}
}
The two shifts are of different character. The one in \Eqnref{VtoC}
involves only $x_f$, which is held fixed by the minimization over $x_0$ and
therefore survives it unchanged, i.e.,
\algn{\eqnlab{VstartoC}
    \ms{V}^{\eta,*}_{t_f}(x_f)=\ms{C}^{\eta,*}_{s\,t_f}(x_f)
    +\frac{\kappa_\text{q}}{4}x_f^2\,.
}
Consequently, such a final shift deforms the rate function by a parabola but
moves neither the kink at $x_f=0$ nor the critical time. The shift
\Eqnref{VtotV}, by contrast, is equivalent to changing the terminal cost $\tilde V$ in the control problem and hence the
transition itself. In particular, \Eqnref{VtotV} lowers the curvature of $\tilde V$ to
$\tilde V''(0) = V_\text{eq}''(0)-\kappa_\text{q}/2$, so that \Eqnref{tcdef}
evaluates differently in the relaxation problem.

For the parabolic class, of which the stochastic work treated in
Ref.~\cite{Mei26b} is the prominent representative, $\kappa_\text{q} = 0$, so that
both shifts in \Eqsref{trafos} disappear and the mapping is direct.

The kink in $\ms{C}^{\eta,*}_{t_f}(u_0)$ at $u_0 = 0$ for $t_f > t_\text{c}^\eta$ therefore
corresponds to a kink in $\ms{V}^{\eta,*}_{t_f}(x_f)$ at $x_f = 0$ for
$t_f > t_\text{c}^{R,\eta}$, where the relaxation critical time $t_\text{c}^{R,\eta}$ follows
from \Eqsref{tcs} with the shifted potential~\Eqnref{VtotV} and by applying the time rescaling of \Eqnref{sfactor},
$t_\text{c}^{R,\eta} = t_\text{c}^{\eta}/s$.

\paragraph*{Explicit verification for each class.}
For the \emph{parabolic class} ($\kappa_\text{q} = 0$, free diffusive
relaxation), the rate function for fixed endpoints is obtained from \Eqnref{C0} with the identification~\eqnref{trafos}
for $\kappa_\text{q}=0$, which gives
\algn{\eqnlab{V0fixed}
    \ms{V}^0_{t_f}(x_f, x_0) = \frac{\gamma(x_f - x_0)^2}{4t_f}
    + V_\text{eq}(x_0)\,.
}
Minimizing over $x_0$ we then find the rate function
\algn{\eqnlab{V0star}
    \ms{V}^{0,*}_{t_f}(x_f) = \min_{x_0}\left\{
    \frac{\gamma(x_f - x_0)^2}{4t_f} + V_\text{eq}(x_0)\right\}\,.
}
The boundary term of \Eqnref{OMsplit} vanishes together with
$\kappa_\text{q}$, so \Eqnref{VtoC} reduces to the plain identification
$V_\text{eq}\leftrightarrow\tilde V$, and the equivalence condition
\Eqnref{equivcond} is satisfied trivially. The time rescaling uses the factor
$s = \xi_R/\xi = 2\kappa\tau_0/\gamma = 2\tau_0/\tau_\text{p}$, which is
the parabolic counterpart of \Eqnref{sfactor}. For the mean-work cost, where $\tau_0 = 2\tau_\text{p}$
(\Secref{canonical}), this gives $s = 4$, so that the mapping
involves the time inversion and rescaling $t\to 4(t_f-t)$ quoted in the companion
Letter~\cite{Mei26b}. Consequently, the kink in $\ms{V}^{0,*}_{t_f}(x_f)$ at $x_f = 0$ forms for $V_\text{eq}=V_\text{DW}$ [cf. \Eqnref{dwell-main}] at the relaxation critical time~\cite{Mei26b}
\algn{\eqnlab{tcR0}
    t_\text{c}^{R,0} = \frac{\gamma x_\text{m}^2}{2V_0}\,.
}

For the \emph{hyperbolic class} ($\kappa_\text{q} > 0$, quench into a harmonic
potential), the rate function for fixed endpoints is obtained from~\Eqnref{Cminus} with the identities~\eqnref{trafos} as
\algn{\eqnlab{Vmfixed}
    \ms{V}^-_{t_f}(x_f, x_0) = \frac{\kappa_\text{q}}{2}\,
    \frac{(x_f - \ee^{-t_f/\tau_\text{p}^R}x_0)^2}
    {1 - \ee^{-2t_f/\tau_\text{p}^R}}
    + V_\text{eq}(x_0)\,,
}
where we have absorbed both additional boundary terms into the first term of \Eqnref{Vmfixed}. Equation~\eqnref{Vmfixed}
is identical to the form found in Ref.~\cite{Mei22a}.
After minimization over $x_0$, the rate function $\ms{V}^{-,*}_{t_f}(x_f)$ then reads
\algn{\eqnlab{Vmstar}
    \ms{V}^{-,*}_{t_f}(x_f) = \min_{x_0}
    \left\{\frac{\kappa_\text{q}}{2}\,
    \frac{(x_f - \ee^{-t_f/\tau_\text{p}^R}x_0)^2}
    {1 - \ee^{-2t_f/\tau_\text{p}^R}} + V_\text{eq}(x_0)\right\}\,.
}
In the hyperbolic class, condition \Eqnref{equivcond} is non-trivial: Using
$|\zeta_R/\xi_R|^{1/2} = \kappa_\text{q}/2$ from \Eqnref{zetxiR} and
$|\zeta/\xi|^{1/2} = \kappa\tau_0/\tau_\text{c}$, \Eqnref{equivcond}
fixes $\kappa_\text{q}$ to
\algn{\eqnlab{kappaq}
	\kappa_\text{q} = 2\kappa\,\frac{\tau_0}{\tau_\text{c}}\,.
}
Hence, to allow for the mapping, the post-quench stiffness $\kappa_\text{q}$
can only agree with the trap stiffness $\kappa$ of the control problem in the special case
$\tau_0/\tau_\text{c} = 1/2$. In general, $\kappa_\text{q}$ and $\kappa$ must differ by twice the ratio $\tau_0/\tau_\text{c}$ of the
control timescales.

The kink in $\ms{V}^{-,*}_{t_f}(x_f)$ forms
at the relaxation critical time
\algn{\eqnlab{tcRm}
    t_\text{c}^{R,-} = \frac{\tau_\text{p}^R}{2}
    \log\!\left(1 + \Gamma_\text{DW}^{-1}\right)\,,
}
with $\Gamma_\text{DW}$ in \Eqnref{Gamdef} formed with the post-quench
stiffness, $\Gamma_\text{DW} = V_0/(\kappa_\text{q}x_\text{m}^2)$. Equation~\eqnref{tcRm}
follows from the control result~\Eqnref{tcminus}  together with the shifted terminal cost of
\Eqnref{VtotV} and scaled with $s$ in \Eqnref{sfactor}.

We note that the shift in \Eqnref{VtotV} not only changes the functional form of \Eqnref{tcRm}, but
it also transforms condition~\eqnref{condition2} into the less restrictive~\eqnref{condition}
for $V_\text{eq}$, i.e., $V_\text{eq}''(0)<0$. Hence, the relaxation transition
exists for every locally concave obstacle, as \Eqnref{tcRm} shows.

At fixed $\tau_\text{p}^R$ and $\Gamma_\text{DW}\to\infty$,
which corresponds to a strong obstacle barrier or a weak harmonic
confinement, $t_\text{c}^{R,-}/\tau_\text{p}^R\to 0$, so the system is in the
ordered phase for essentially all $t_f$. If instead $\kappa_\text{q}\to0$ at
fixed $V_0$ and $x_\text{m}$, then $\tau_\text{p}^R=\gamma/\kappa_\text{q}$ diverges together with
$\Gamma_\text{DW}$ and the product stays finite,
$t_\text{c}^{R,-}\to\gamma x_\text{m}^2/(2V_0) = t_\text{c}^{R,0}$ of \Eqnref{tcR0},
as expected from the convergence of the hyperbolic to the parabolic class.

The mapping \Eqsref{trafos}--\eqnref{VstartoC} is the central result of
this section. It shows that measuring the optimal control cost
$\ms{C}^{\eta,*}_{t_f}(u_0)$ in an optimal control experiment gives direct access to the
rate function $\ms{V}^{\eta,*}_{t_f}(x_f)$ of the associated relaxation process,
and that the control transition is an experimentally accessible signature of the FTDPT.
This mapping provides a practical route to measuring dynamical phase
transitions without the exponentially large sample sizes that direct
observation of rare relaxation events would require.

% -------------------------------------------------------
\subsection{Key differences between control and relaxation}
\seclab{differences}
% -------------------------------------------------------

The mapping \Eqsref{trafos}--\eqnref{VstartoC} establishes a correspondence between the optimal control cost and the
large-deviation rate function. However, the two quantities have different physical interpretations and experimental signatures. We now
discuss the principal differences between the control transition and
the FTDPT, which together explain why the control experiment provides a
more accessible route to the same physics.

\paragraph*{Nature of cost.}
The optimal control cost $\ms{C}^{\eta,*}_{t_f}(u_0)$ is a statistical
\emph{average} over an ensemble of controlled trajectories. For a given
protocol duration $t_f$ and initial position $u_0$, it equals the cost averaged over many realizations of the thermal
noise. It is therefore a smooth, well-defined quantity at any finite noise
level, requiring only $O(1)$ sample trajectories to estimate to a given
relative precision.

The rate function $\ms{V}^{\eta,*}_{t_f}(x_f)$, by contrast, characterizes an
\emph{exponentially rare} event: the probability of finding the
relaxing particle at position $x_f$ at time $t_f$ is suppressed as
$P(x_f, t_f)\propto\exp[-\ms{V}^{\eta,*}_{t_f}(x_f)/(\kb T)]$. Away from the
regions where $\ms{V}^{\eta,*}_{t_f}$ is small, estimating $P(x_f, t_f)$ to a
given relative precision requires a number of sample trajectories that
grows as $\exp[\ms{V}^{\eta,*}_{t_f}(x_f)/(\kb T)]$. For typical experimental
parameters with $\ms{V}^{\eta,*}_{t_f}/(\kb T)\sim 10$--$40$, this corresponds
to at least $10^4$ trajectories, which would all need to be initially equilibrated in the potential $V_\text{eq}(x)$. This represents a sampling challenge that makes
direct observation of the FTDPT prohibitively expensive without special
techniques. The control experiment bypasses this challenge, since
it accesses equivalent information through direct averages over a manageable number of
\emph{controlled} trajectories, where the protocol actively guides the
particle to the relevant region of phase space.

\paragraph*{Sharpness and weak-noise limit.}
The kink in $\ms{C}^{\eta,*}_{t_f}(u_0)$ at $u_0 = 0$ for $t_f > t_\text{c}^\eta$ is
\emph{perfectly sharp} at any finite noise level in the ideal limit of infinite statistics. In practice, the kink is rounded by finite sampling and measurement error. This follows directly
from the structure of the optimal control problem: the kink arises from a
discontinuous switch of the minimizer $u_f^*(u_0, t_f)$ at $u_0 = 0$, which
is a property of the deterministic cost function $\ms{C}^\eta_{t_f}(u_f, u_0)$
and is unaffected by the magnitude of thermal fluctuations. In
particular, the noise enters the control problem only through the
noise-averaged penalty $\tilde V(u_f)$, and the kink exists whenever
the condition~\eqnref{condition} or \eqnref{condition2} is fulfilled, regardless of the noise level.

The kink in $\ms{V}^{\eta,*}_{t_f}(x_f)$, by contrast, is sharp only in the
strict weak-noise limit $\kb T/V_0\to 0$, with the extra condition $\kappa_\text{q}x_\text{m}^2\sim V_0$
for relaxation into a harmonic potential. At any finite $\kb T/V_0$, the large-deviation
approximation~\eqnref{LDform} receives subleading corrections that smooth
out the kink.
The FTDPT is therefore a sharp transition only in the thermodynamic sense
of a limit, analogous to an equilibrium phase transition that becomes sharp
only as $N\to\infty$. For the experiments described in \Secref{experiments},
the finite-$\kb T/V_0$ smoothing is noticeable in the rate function data and
must be taken into account when comparing with theory.

\paragraph*{Critical fluctuations.}
At the critical time $t_f = t_\text{c}^\eta$, the control transition shows no critical
fluctuations. The order parameter $u_f^*(t_f)$ is continuous at
$t_\text{c}^\eta$, growing as $u_f^*\propto(t_f-t_\text{c}^\eta)^{1/2}$
[\Eqnref{mfexp}], so that its slope diverges while the transition itself
remains continuous. In our optimal control experiment, the variance of the measured work around its mean remains finite and
unremarkable at $t_\text{c}^\eta$. This is because $u_f^*$ is a deterministic
minimizer of a deterministic cost function.

The FTDPT, by contrast, exhibits genuine critical fluctuations of the
order parameter $x_0^*(x_f = 0, t_f)$ as $t_f\to t_\text{c}^{R,\eta}$. At small but finite noise, the
susceptibility
\algn{\eqnlab{suscept}
    \chi_{t_f} = \frac{V_0}{\kb T\,x_\text{m}^{2}}\left[
    \langle x_0^2\,|\,x_f = 0,\,t_f\rangle
    - \langle |x_0|\,|\,x_f = 0,\,t_f\rangle^2\right]
}
develops a peak around $t_f\approx t_\text{c}^{R,\eta}$ whose height grows
as $(V_0/\kb T)^{1/2}$ as $\kb T/V_0\to0$. Away from the peak, and in that
same limit, $\chi_{t_f}$ behaves as  $\propto|t_f-t_\text{c}^{R,\eta}|^{-1}$ on both
sides of the transition. Together
with $\beta_\text{MF}=1/2$ this identifies the mean-field exponent
$\gamma_\text{MF}=1$~\cite{Vad24}.
When multiplied by $(\kb T/V_0)^{1/2}$ and plotted against the scaling
variable $\theta$ of \Eqnref{theta}, the susceptibility curves for different
$\kb T/V_0$ collapse onto a universal scaling function,
providing a clear experimental signature of the FTDPT that is absent in
the control experiment. We observe this peak and the associated
scaling collapse in the free-relaxation experiment described in Ref.~\cite{Mei26b} and the harmonic-relaxation experiments
described in \Secref{experiments}. Despite the exponential sampling cost of observing the FTDPT directly, the
statistical challenges described above can be partially overcome by the importance-sampling
technique described in \Secref{experiments}. The key observation is that
the rate function $\ms{V}^{\eta,*}_{t_f}(x_f)$ can be reconstructed from an
ensemble of \emph{freely diffusing} or \emph{relaxing} trajectories,
with no reference to the initial Boltzmann distribution $P_\text{eq}(x_0)\propto
\ee^{-V_\text{eq}(x_0)/(\kb T)}$. The initial state is subsequently implemented
in a post-processing step where the trajectories are \emph{reweighted} so that their initial distribution
matches the equilibrium Boltzmann weight $P_\text{eq}(x_0)$.
The result is an estimate of $\ms{V}^{\eta,*}_{t_f}(x_f)$ from $O(10^5)$
measured \emph{trajectory intervals}, compared to at least $O(10^4)$ separately \emph{equilibrated
trajectories}, each costing a barrier-crossing time, that direct sampling
would require at the parameter values used in the experiments.

Table~\ref{tab:differences} collects the key differences between the
control transition and the FTDPT discussed above.

\begin{table*}[t]
\caption{Key differences between the finite-time control transition and
the finite-time dynamical phase transition (FTDPT) in non-equilibrium
relaxation, for the parabolic and hyperbolic classes.}
\label{tab:differences}
\begin{ruledtabular}
\begin{tabular}{lll}
\textbf{Property} & \textbf{Control transition} & \textbf{FTDPT} \\
\hline
Nature of cost         & Statistical average         & Exponentially rare probability \\
Sampling cost          & $O(1)$ trajectories         & $O[\exp(\ms{V}_{t_f}/\kb T)]$ trajectories \\
Sharpness of kink      & Sharp at any noise level    & Sharp only in weak-noise limit \\
Critical fluctuations  & Absent                      & Present with mean-field exponents \\
Susceptibility at $t_\text{c}$& Finite                      & Peak $\propto(V_0/\kb T)^{1/2}$, $\propto|t_f-t_\text{c}|^{-1}$ off peak \\
Experimental method    & Controlled trajectories     & Free/relaxing trajectories + reweighting \\
\end{tabular}
\end{ruledtabular}
\end{table*}

% =========================================================
\section{Experiments}
\seclab{experiments}
% =========================================================

We have designed three complementary experiments to probe the finite-time
control transition and the associated FTDPT identified in the preceding
sections. The first experiment directly measures a control transition in
the parabolic class. The second and third experiments probe FTDPTs from the parabolic and hyperbolic classes, respectively. The parabolic control transition and free-relaxation FTDPT are reported in full in the companion Letter~\cite{Mei26b} and summarized only briefly below, while the hyperbolic-class harmonic relaxation experiment is presented here in more detail. All three experiments
use the same colloidal system manipulated by optical tweezers, described
below.

% -------------------------------------------------------
\subsection{Experimental setup}
\seclab{setup}
% -------------------------------------------------------

\paragraph*{Colloidal suspension.}
Our experimental system consists of a single spherical silica particle
(diameter $2.73\pm 0.12\,\unit{\micro\meter}$, microParticles GmbH)
suspended in a 1:1 volume fraction mixture of water and glycerol. The
mixture is contained in a glass capillary (inner diameter
$100\,\unit{\micro\meter}$, width $1\,\unit{\milli\meter}$; CM Scientific)
sealed with wax and epoxy resin. We determined the friction coefficient $\gamma$ for each from fits to
the equilibrium mean-squared displacement: $\gamma \approx 0.11\,\unit{\micro N s/m}$ for the optimal control experiment,
$0.12\,\unit{\micro N s/m}$ for the harmonic relaxation experiment, and
$0.18\,\unit{\micro N s/m}$ for the free relaxation experiment.
The particle is trapped in the midplane
of the capillary to avoid hydrodynamic interactions with the walls.
Samples are allowed to equilibrate for at least 60 minutes before data
acquisition.

\paragraph*{Optical tweezers.}
The particle is manipulated by optical tweezers formed by a focused
532~nm laser beam (Coherent Verdi V2) directed through an oil-immersion
microscope objective (Olympus MPLAPON-Oil, 100$\times$, NA$=1.45$). The
laser creates a trapping potential that is very accurately described by a harmonic potential for the displacements from the trap center relevant in our experiments 
\algn{\eqnlab{trap}
    U(x,\lambda) = \frac{\kappa}{2}(x - \lambda)^2\,,
}
where $\lambda$ is the time-dependent trap center and $\kappa$ is the trap
stiffness. Across different experimental runs and consistent with the
5\% polydispersity of the particles, the trap stiffness ranged over
$\kappa = (0.51\text{--}0.62)\pm 0.05\,\unit{\micro N/m}$, giving
relaxation times $\tau_\text{p} = \gamma/\kappa = (0.172\text{--}0.216)
\pm 0.02\,\unit{s}$.

\paragraph*{Trap control.}
Dynamic positioning of the trap center $\lambda(t)$ is achieved in two
complementary ways, depending on the experiment. For the control
experiment (\Secref{controlexp}), the laser beam is deflected by an
acousto-optical deflector (AOD, AA Opto-Electronic DTSXY-400) with a
spatial accuracy of 10~nm and a repetition rate of $10^{-4}\,\unit{s}$.
This allows arbitrary time-dependent protocols to be implemented with
high precision. For the harmonic relaxation experiment (\Secref{harmexp}),
where the trap must be repositioned quasi-instantaneously between fixed
locations, a piezo-driven mirror is used instead, which achieves faster
repositioning with lower positional noise.

\paragraph*{AOD calibration.}
The AOD exhibits a frequency-dependent diffraction efficiency that induces
spatial variations in the trap stiffness across the field of view. We
calibrate these variations by sampling a spatial grid and assigning
time-averaged particle positions to the corresponding laser focus
locations. The trap stiffness $\kappa_i$ is measured at each grid point
and normalized to its mean value $\langle\kappa_i\rangle$ in each spatial
direction. A correction factor
\algn{\eqnlab{corr}
    c(f_x,f_y) = \frac{\kappa_x/\langle\kappa_x\rangle
    + \kappa_y/\langle\kappa_y\rangle}{2}
}
is fitted by a third-order polynomial $C(f_x,f_y)$ over the field of
view. By construction, $C\approx 1$, on average, and the applied AOD
amplitude is rescaled as
\algn{\eqnlab{corramp}
    A_\text{apply} = \frac{A_\text{set}}{0.5 + 0.5\cdot C(f_x,f_y)}\,,
}
which returns $A_\text{apply} = A_\text{set}$ when $C = 1$. This
procedure reduces spatial variations in trap stiffness from $\pm 20\%$
to $\pm 5\%$.

\paragraph*{Particle tracking.}
Particle positions are tracked using digital video microscopy. Videos are recorded at 250~Hz
for the control experiment and at 400~Hz for the relaxation experiments, with a spatial resolution of 10\,$\unit{\nano\meter}$. The sample temperature is maintained at $25\pm 0.05\,\unit{\celsius}$ by resistive heating applied to both the sample stage and the objective (Okolab H401-T-Penny).

% -------------------------------------------------------
\subsection{Optimal control experiment: parabolic class}
\seclab{controlexp}
% -------------------------------------------------------

\paragraph*{Setup and protocol.}
We consider the mean stochastic work~\eqnref{Wstoch}
as the cost functional, placing the problem in the parabolic class, for which the optimal protocol is given in \Eqnref{lam0}. 
The obstacle is modeled by the symmetric double-well potential~\eqnref{dwell-main}. The potential minima correspond to the openings of a soft double slit, and the barrier of height $V_0/4$ at $x = 0$ corresponds to the obstacle itself. The
noise-averaged penalty $\tilde V(u_f)$ is derived in Appendix~\secref{Apotential}.
For the experimental parameters, the noise level $\eps = \kb T/(\kappa x_\text{m}^2)
\approx7\times10^{-3}$ coincides with $\kb T/V_0$ because $V_0 = \kappa x_\text{m}^2$.
Hence, $\eps$ is small enough that $\tilde V(u_f)\approx V(u_f)$ to
good approximation.

For each initial position $u_0$ and protocol duration $t_f$, we implement the
optimal protocol \Eqnref{lam0} by moving the AOD-controlled trap center
$\lambda(t)$, and evaluate the stochastic work from the measured trajectories
via~\cite{Cro98}
\algn{\eqnlab{Wdisc}
    W = \sum_{i=1}^N\left\{ U[x(t_i),\lambda(t_{i+1})] - U[x(t_i),\lambda(t_i)]
    \right\}
    \,,
}
where the sum runs over $N$ intervals of duration $t_f/(N-1) = 0.004\,\unit{s}$,
set by the $250\,\unit{Hz}$ acquisition rate of the camera in the control experiment.
Forward and backward implementations are averaged to remove trap drift. More
procedural details are given in Appendix~\secref{Aexpcontrol}.

\paragraph*{Measured control transition.}
These measurements reproduce the finite-time control transition of the
parabolic class. For $u_0 = 0$, the optimal cost
$\ms{W}^*_{t_f} = \langle W_{t_f}\rangle$ is flat at the obstacle penalty
$\tilde V(0) = V_0/4$ for $t_f < t_\text{c}$ and decreases for $t_f > t_\text{c}$ as the
optimal protocol steers the particle toward a minimum of $\tilde V$.
As a function of $u_0$, $W^*_{t_f}$ is smooth below $t_\text{c}$ and
develops a sharp kink at $u_0 = 0$ above $t_\text{c}$, reflecting the discontinuous
switch of $u_f^*(u_0, t_f)$ (\Secref{landau}). The extracted order parameter
$|u_f^*|$ vanishes below $t_\text{c}$ and grows as $(t_f - t_\text{c})^{1/2}$ above it,
confirming the mean-field exponent. The measured optimal cost, order
parameter, optimal protocols, and sample trajectories are reported in the
companion Letter~\cite{Mei26b}.

% -------------------------------------------------------
\subsection{Free relaxation experiment: parabolic class FTDPT}
\seclab{freeexp}
% -------------------------------------------------------
\begin{figure*}[t]
    \centering
    \includegraphics[width=\linewidth]{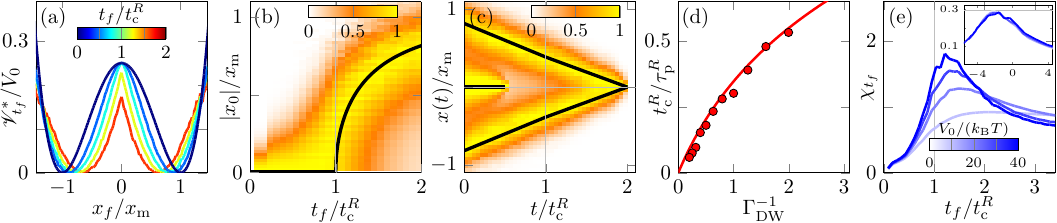}
    \caption{Results of the harmonic relaxation experiment for $V_0\approx 20\,\kb T$.
    (a) Measured rate function $\ms{V}^{-,*}_{t_f}(x_f)$ developing a kink at $x_f = 0$ for
    $t_f > t_\text{c}^{R,-}$. (b) Order-parameter density $P_\text{op}(|x_0|, t_f)$ as a heat
    map over $t_f$ and $|x_0|$. The solid line shows the theoretical
    most-likely initial position $x_0^*(t_f)$. (c) Optimal-fluctuation
    trajectory density for two protocol durations, one below and one above
    $t_\text{c}^{R,-}$. Solid lines are the theoretical optimal fluctuations given in the text.
    (d) Critical time $t_\text{c}^{R,-}$ as a function of $\Gamma_\text{DW}^{-1}$ from experiment (markers) and theory [\Eqnref{tcRm}, line].
    (e) Susceptibility $\chi_{t_f}$ of \Eqnref{suscept} at
    $\Gamma_\text{DW} = 1$ for
    $V_0/\kb T = 10,\,20,\,30,\,40$ (color bar). Inset: Same data as in (e), showing
    $(\kb T/V_0)^{1/2}\chi_{t_f}$ as a function of the scaling variable
    $\theta^-$ [\Eqnref{theta}] showing the collapse onto a scaling function. The light grey line marks its
    maximum $\approx0.2973$.}
    \figlab{relaxexp}
\end{figure*}
\paragraph*{Setup and importance sampling.}
In the second experiment, we probe the FTDPT in the non-equilibrium
relaxation of a freely diffusing particle initially at equilibrium in the
double-well potential \Eqnref{dwell-main}. The rate function
$\ms{V}^{0,*}_{t_f}(x_f)$ is predicted to develop a kink at $x_f = 0$ for
$t_f > t_\text{c}^{R,0}$, given by \Eqnref{tcR0}. The large-deviation
parameter controlling the sharpness of the transition is $\kb T/V_0\approx
0.025$. This value is small enough that the kink is clearly visible but large enough
to ensure sufficient statistics.

Direct observation of the FTDPT would require
at least $O[\exp(\ms{V}^{0,*}_{t_f}/\kb T)] \approx 10^4$ sample trajectories for
the parameters used here, each of which would have to be equilibrated afresh in the
initial potential $V_\text{eq}(x)$. To circumvent this, we employ the
importance-sampling technique mentioned briefly in~\Secref{differences}. We record
approximately $2\times 10^3$ trajectories of a freely diffusing particle
(trap switched off) each of duration $t_\text{off} = 0.5\,\unit{s}$, during intervals
between trap-on periods of equal length. With the $x_\text{m} = 0.2\,\unit{\micro\meter}$ and
$V_0 = 40\,\kb T$ used in the reweighting, this corresponds to
$t_\text{off}\approx 22\,t_\text{c}^{R,0}$, so that the snippet durations
are long enough to observe the transition. We then exploit the translational invariance of free diffusion to
shift all trajectories to a common starting point $x_0=0$, and use the Markov
property to split each long trajectory into multiple overlapping shorter
snippets of the required duration $t_f$, amplifying the effective sample
size to $\approx 10^6$ snippets. The resulting empirical distribution of
initial positions $x_0$ is reweighted by the normalized Boltzmann factor
$\ee^{-V(x_0)/(\kb T)}/Z$ to simulate sampling from the equilibrium
distribution in the double-well. From the reweighted ensemble we compute
the rate function as
\algn{\eqnlab{Vfromdata}
    \ms{V}^{0,*}_{t_f}(x_f) \approx -\kb T\log P_\text{rw}(x_f, t_f)\,,
}
where $P_\text{rw}(x_f, t_f)$ is the reweighted probability density of the
final position. Full details of the reweighting procedure are given in
Appendix~\secref{Aexpfreerelax}.

\paragraph*{Measured FTDPT.}
The reweighted rate function $\ms{V}^{0,*}_{t_f}(x_f)$ develops a kink at
$x_f = 0$ at $t_f = t_\text{c}^{R,0}$, and the order-parameter density
$P_\text{op}(|x_0|, t_f)=P(|x_0|,0|0,t_f)$ reorganizes from $x_0\approx 0$ to $|x_0|\approx x_\text{m}$ across
the transition, confirming the FTDPT in the parabolic class. The
susceptibility $\chi_{t_f}$ of \Eqnref{suscept} develops a pronounced peak
whose finite-$\kb T/V_0$ scaling collapse demonstrates mean-field critical
exponents, a critical behavior that is absent from the control experiment.
The measured rate function, order-parameter density, trajectory densities,
and susceptibility collapse are reported in the companion Letter~\cite{Mei26b}.

% -------------------------------------------------------
\subsection{Harmonic relaxation experiment: hyperbolic class FTDPT}
\seclab{harmexp}
% -------------------------------------------------------
%
\paragraph*{Setup.}
The third experiment probes an FTDPT in the hyperbolic class, by measuring the relaxation of a particle into a harmonic potential
with stiffness $\kappa_\text{q} > 0$ after equilibration in the double-well
\Eqnref{dwell-main}. Here $\kappa_\text{q}$ is the post-quench stiffness of
\Secref{mapdetail}, which in this experiment is realized by the optical
tweezers. The large-deviation parameter for this
experiment is $\kb T/V_0\approx
0.05$. In contrast to the free-relaxation experiment,
the post-quench dynamics are not translationally invariant, because the
harmonic potential breaks translational symmetry. This modifies the
importance-sampling procedure and requires a different experimental
protocol for generating the trajectory ensemble.

Rather than switching the trap fully off, we implement the quench by
instantaneously repositioning the trap center $\lambda$ between two fixed
locations using a piezo-driven mirror. In the time intervals of duration
$t_\text{rel}\approx 0.5\,\unit{s}$ between consecutive jumps, the
particle relaxes in the harmonic potential toward its new equilibrium at
$\lambda$. We record the particle position throughout these relaxation
intervals and divide each trajectory into overlapping snippets of the
required duration $t_f$ using the Markov property. The reweighting procedure
is then applied as described in \Secref{freeexp}, adapted to account for
the absence of translational invariance (see Appendix~\secref{Aexpharmrelax} for
details). Approximately $1\times 10^7$ overlapping trajectory snippets are
obtained from $1.2\times 10^5$ relaxation intervals.

\paragraph*{Measured rate function and order parameter.}
Figure~\ref{fig:relaxexp}(a) shows the measured rate function
$\ms{V}^{-,*}_{t_f}(x_f)$ for several values of $t_f$, together with the
theoretical prediction from \Eqnref{Vmstar}. For $t_f < t_\text{c}^{R,-}$ the
rate function has a smooth double-well shape. At $t_f = t_\text{c}^{R,-}$ it
develops a kink at $x_f = 0$ that deepens for $t_f > t_\text{c}^{R,-}$,
confirming the FTDPT in the hyperbolic class. A key qualitative difference
from the parabolic case is that the minima of $\ms{V}^{-,*}_{t_f}$ move toward
$x_f = 0$ as $t_f$ increases, reflecting the restoring force of the harmonic
potential that pulls the particle back to the trap center. At long times
$\ms{V}^{-,*}_{t_f}$ approaches the equilibrium rate function of the harmonic
potential, which has a single minimum at $x_f = 0$.

Figure~\ref{fig:relaxexp}(b) shows the order parameter density
$P_\text{op}(|x_0|, t_f)=P(|x_0|,0|0,t_f)$ as a heat map, together with the theoretical prediction
for the most likely initial position $|x_0^*|(t_f)$. The transition at
$t_f = t_\text{c}^{R,-}$ is visible as a shift of the density from near $x_0 = 0$
to near $|x_0| = x_\text{m}$.

\paragraph*{Trajectory density and optimal fluctuations.}
Figure~\ref{fig:relaxexp}(c) shows the optimal-fluctuation trajectory density
conditioned on reaching $x_f = 0$ at time $t_f$, for one duration below and
one above $t_\text{c}^{R,-}$. For $t_f < t_\text{c}^{R,-}$ the dominant fluctuation
starts at $x_0\approx 0$ and remains near zero throughout, in analogy
with the parabolic case reported in the Letter~\cite{Mei26b}, and in agreement with the behavior of the optimal
trajectories in the control problems in \Secref{transition}. For $t_f > t_\text{c}^{R,-}$, however, the optimal
fluctuations are no longer straight lines, but start near $\pm x_\text{m}$ and follow hyperbolic functions toward $x_f=0$. The solid lines in \Figref{relaxexp}(c) show the optimal fluctuation $x^*(t) = \{\sinh[(t_f-t)/\tau_\text{p}^R]\,x_0
+ \sinh(t/\tau_\text{p}^R)\,x_f\}/\sinh(t_f/\tau_\text{p}^R)$ obtained from $u^*(t)$ in \Eqnref{uhyp} using the mapping in \Secref{mapdetail}.

\paragraph*{Critical time as function of $\Gamma_\text{DW}$.}
A distinctive feature of the hyperbolic class FTDPT is that the critical
time $t_\text{c}^{R,-}$ depends on the dimensionless parameter
$\Gamma_\text{DW} = V_0/(\kappa_\text{q} x_\text{m}^2)$, the ratio of the potential barrier height to the
harmonic energy scale, as given by \Eqnref{tcRm}. Exploiting the fact
that the reweighting procedure is equivalent to virtually adjusting $x_\text{m}$
(and hence $\Gamma_\text{DW}$) without changing the measured trajectories, we can
extract $t_\text{c}^{R,-}(\Gamma_\text{DW})$ from a single set of experimental trajectories
by applying the reweighting with different values of $x_\text{m}$. Figure~\ref{fig:relaxexp}(d) shows that $t_\text{c}^{R,-}$ increases monotonically
with increasing $\Gamma_\text{DW}^{-1} = \kappa_\text{q} x_\text{m}^2/V_0$, in excellent agreement
with the theoretical prediction \Eqnref{tcRm}. In the limit $\Gamma_\text{DW}\to\infty$
(strong barrier relative to the harmonic energy scale, or equivalently
weak harmonic confinement), $t_\text{c}^{R,-}\!\!\to 0$, so the system is in the
ordered phase for all protocol durations. In the limit $\kappa_\text{q}\to 0$,
the critical time $t_\text{c}^{R,-}$ approaches $t_\text{c}^{R,0}$, as
discussed previously.

\paragraph*{Susceptibility.}
Figure~\ref{fig:relaxexp}(e) shows the susceptibility $\chi_{t_f}$ of
\Eqnref{suscept}, measured at $\Gamma_\text{DW} = 1$ for four values of
$\kb T/V_0\ll 1$
obtained from the same trajectory ensemble by reweighting with different
$x_\text{m}$. The peak grows as $(V_0/\kb T)^{1/2}$ and migrates towards
$t_\text{c}^{R,-}$ as $\kb T/V_0$ decreases. The finiteness of $\chi_{t_f}$ corresponds to the finite-noise
rounding of the kink in $\ms{V}^{-,*}_{t_f}(x_f)$ at $x_f=0$ and $t_f=t_\text{c}^{R,-}$.

The collapse follows from the structure of the conditioned distribution.
Conditioned on $x_f = 0$, both \Eqnref{V0fixed} and \Eqnref{Vmfixed} reduce,
in $y = x_0/x_\text{m}$, to the same constrained free energy
\algn{\eqnlab{redfree}
	F^\eta(y,t_f) = \frac{{\tilde\Phi}^{\eta}_R(t_f)}{2}y^2 + \frac{1}{4}\left(y^2-1\right)^2\,,
}
with 	$P(y,t_f)\propto\exp[-V_0F(y,t_f)/\kb T]$, differing only through the dimensionless curvature ${\tilde\Phi}^{\eta}_R(t_f)$, which equals
${\tilde\Phi}^{0}_R(t_f)=t_\text{c}^{R,0}/t_f$ for the parabolic class and
\algn{\eqnlab{phiR}
	{\tilde\Phi}^{-}_R(t_f) = \frac1{\Gamma_\text{DW}(\ee^{2t_f/\tau_\text{p}^R}-1)}\,,
}
for the hyperbolic class, while the transition is at ${\tilde\Phi}^{\eta}_R(t^{R,\eta}_c) = 1$ in both cases. Substituting
$y = (4\kb T/V_0)^{1/4}z$ gives $V_0F/\kb T = \theta z^2 + z^4$ with
\algn{\eqnlab{theta}
    \theta^\eta(t_f) = [{\tilde\Phi}^{\eta}_R(t_f)-1]\left(\frac{V_0}{\kb T}\right)^{1/2}\,,
}
so that $(\kb T/V_0)^{1/2}\chi_{t_f} = 2\,\text{var}(z)$ is a universal
function of $\theta^\eta$ alone, independent of $\kb T/V_0$ and of $\Gamma_\text{DW}$. The inset of Fig.~\ref{fig:relaxexp}(e) shows $(\kb T/V_0)^{1/2}\chi_{t_f}$
as a function of $\theta^-$, confirming the collapse for the hyperbolic class.
Because $\theta^\eta$ absorbs all parameter dependence, the rescaled height of $(\kb T/V_0)^{1/2}\chi_{t_f}$
attains a universal maximum $\approx 0.2973$, which can be obtained explicitly, without any adjustable parameter. The measured
peak heights agree with this value to within a few percent, see the light grey line in the inset of~\Figref{relaxexp}(e).

% =========================================================
\section{Discussion and Outlook}
\seclab{discussion}
% =========================================================

\paragraph*{Summary of classification.}
The central result of this paper is the classification of all linearized
optimal control problems of the form~\eqnref{meandyn}--\eqnref{cost} into
three disjoint canonical equivalence classes, distinguished by the sign of
\algn{
    \zeta = -2(C_{11} + 2C_{12} + C_{22})\,,
}
or equivalently by the sign of $\det\Hmat = \zeta\xi$. The classification is exhaustive
within the class of quadratic cost functionals and linear dynamics, and it
is canonical in the sense that any two problems with the same sign of
$\det\Hmat$ can be related by a canonical transformation
of the control Hamiltonian. The three classes
have qualitatively distinct optimal protocols, cost functions, and critical
times, but share the universal features of the transition described above.
The boundary jumps of the optimal protocol, by contrast, are not
universal. They originate from control-dependent boundary terms that naturally occur due to $\dot\lambda$-terms of work-like cost functionals~\cite{Sch07}.
In the absence of such boundary terms, the endpoints of the protocol are undefined and the jumps can be chosen to disappear. The parabolic class ($\zeta = 0$) contains
the mean-work problem studied experimentally in~\cite{Mei26b} and in
\Secref{controlexp}. The hyperbolic class ($\zeta < 0$) is realized by
the quadratic control cost \Eqnref{Cquad} and connects to the FTDPT in
harmonic relaxation studied in \Secref{harmexp}. By contrast, the elliptic class
($\zeta > 0$) is a genuine control-theoretic phenomenon with no direct relaxation
counterpart. In particular, beyond the time $t_\text{in} = \pi\tau_\text{c}$ the
linearized elliptic problem is unstable (\Secref{hamclass}), which is explained
by the fact that for $\zeta>0$, mutual translations of $\lambda$ and $u$ are rewarded,
leading to optimal protocols that harvest large negative costs.

\paragraph*{Generality of linearization.}
The linearized framework derived in \Secref{model} applies whenever two
conditions hold: the existence of a zero-control fixed point $f(x^*,
\lambda^*) = 0$ of the noise-free dynamics, and the smallness of deviations
from that fixed point during the optimal protocol. The second condition is
self-consistent near the transition at $t_f\approx t_\text{c}^\eta$, where the
optimal final position $u_f^*$ is small by definition. For $t_f$ substantially
larger than $t_\text{c}^\eta$, the optimal $u_f^*$ may become large enough that the
linearized framework is only quantitatively approximate. Crucially, this does
not change the classification or the existence of the transition, but affects
only the quantitative form of the optimal cost and the precise value of the
order parameter away from $t_\text{c}$. Similarly, weak symmetry-breaking terms in
$\tilde V$ are consistent with the linearization [see Appendix~\secref{Aasym}], as long as they
do not shift the critical point out of the linear regime.

\paragraph*{Extension to higher dimensions.}
Throughout this paper, we have restricted to one spatial dimension. In $d$
dimensions the phase-space vector $\wvec$ lives in $\mathbb{R}^{2d}$, and the
canonical transformation procedure of \Secref{canonical} generalizes via the
theory of symplectic normal forms for quadratic
Hamiltonians~\cite{Wil36,Arn01}, which classifies the problems by the Jordan form of
$\Jmat\Hmat$.

The simplest generalization is the isotropic case, in which both the
dynamics and the obstacle are invariant under rotations. Then $\Hmat$ is a $2\times2$ block matrix, where each
block is proportional to the identity matrix. The eigenvalues of $\Jmat\Hmat$ are then the $d$-fold degenerate
pair $\pm(-\zeta\xi)^{1/2}$, and the cost depends on $\ve{u}_f$ only through $|\ve{u}_f|$,
so that the radial problem reduces to the one-dimensional case of~\Secref{hamclass}, with
the same $\Phi^\eta(t_f)$ and the same forms for the critical times~\eqnref{tcs}. Two things change, however: First, the
class is labeled by $\text{sgn}(\zeta)$ rather than by $\text{sgn}(\det\Hmat)$, since
$\det\Hmat = (\zeta\xi)^d$ is positive for even $d$ irrespective of the class. Second, the
transition breaks the continuous symmetry $O(d)\to O(d-1)$ instead of $\mathbb{Z}_2$, so
that the optimal final position $|\ve u^*_f|$ is degenerate on a $(d-1)$-dimensional sphere. The mean-field exponents are
unchanged, while the noise correction to $\tilde V$ and the universal amplitude of
$\chi_{t_f}$ acquire $d$-dependences.

For anisotropic $d > 1$, mixed classes can arise in which some
eigenvalue pairs are real and others imaginary, so that hyperbolic and
elliptic behavior coexist in different directions. Consequently, the classification is then no longer
exhausted by a single sign $\eta$, and degenerate eigenvalues require the full Jordan
structure of $\Jmat\Hmat$~\cite{Arn01}.

\paragraph*{Non-Gaussian noise, memory, and active baths.}
The framework of this paper assumes Gaussian white noise and a Markovian evolution, which leads to
the additive-noise Langevin equation \Eqnref{LEQlin} after linearization.
Many experimentally relevant systems involve non-Gaussian or colored noise,
such as active Brownian particles in bacterial baths~\cite{Aru16}, or
particles in viscoelastic media with exponentially decaying memory
kernels~\cite{Loo24}. The classification of such problems
requires an extension of the canonical transformation framework, which is a possible direction for future
work.

\paragraph*{Optimal control in rare-event physics.}
A broader implication of the mapping established in \Secref{mapping} is
that optimal control experiments provide a general and practical route to
rare-event physics in mesoscopic systems~\cite{Fle06,Che15b}. The rate function $\ms{V}^*_{t_f}$
governing rare relaxation trajectories can be accessed through
averages over controlled trajectories, without the exponential sampling
cost that makes direct observation intractable. A natural next step is interacting systems, for
which the cost functional couples the control problems of the individual
particles~\cite{Mon26}. At mean-field level, the effective single-particle problem is again
of the form studied here, so that the classification should survive close to the critical point.
Extensions to quantum systems, where the role of thermal noise is played by
quantum fluctuations~\cite{Hey13,Asr26}, are a further direction.
% =========================================================
\section{Conclusion}
\seclab{conclusion}
% =========================================================

We have developed a theoretical framework for finite-time
transitions in optimal control of stochastic systems in structured
environments, and validated it experimentally using optically trapped
colloidal particles. Starting from a nonlinear overdamped Langevin
equation, we derived a universal linearized control problem by expanding
about the zero-control fixed point of the noise-free dynamics, and showed
that the resulting optimal control problems fall into three canonical
equivalence classes distinguished by the sign of $\det\Hmat$. For each
class, we derived the optimal protocols, cost functions, and critical times
analytically, and established the Landau analogy with a continuous phase
transition, including the mean-field critical exponent $\beta_\text{MF} =
1/2$ for the order parameter. We showed that for the parabolic and
hyperbolic classes the optimal control cost maps exactly onto the
large-deviation rate function governing non-equilibrium relaxation after
a potential quench, via an exchange of boundary conditions, a time
reversal and rescaling, and a harmonic shift of the terminal cost.
The elliptic class, by contrast, has no relaxation counterpart and, being
unstable beyond a finite time $t_\text{in}$, no long-time regime. Three
experiments using optically trapped colloidal particles confirmed the
control transition in the parabolic class, the associated FTDPT in free
relaxation, and the FTDPT in harmonic relaxation (hyperbolic class),
with the experimental rate functions and order parameter densities in
good agreement with theory. Nevertheless, the two transitions are not
equivalent physical phenomena, because only the relaxation transition is
accompanied by critical fluctuations of the order parameter, whose
susceptibility we measured and found to collapse onto a universal scaling
function. The importance-sampling technique employed in
the relaxation experiments demonstrates that FTDPTs can be probed
in experiment with a manageable number of trajectories. Our results
establish finite-time control transitions as a generic feature of optimal
control in structured mesoscopic environments, and provide a complete
classification of their possible types in one spatial dimension.
% =========================================================
\begin{acknowledgments}
	JM thanks John Bechhoefer for pointing out the relevance of Refs.~\cite{Kap05a,Kap05b}. CB acknowledges funding by the Deutsche Forschungsgemeinschaft, SFB 1432 (425217212) and the  ERC AdG.Grant No.101141477.
\end{acknowledgments}
\section*{Data availability}
The data that support the findings of this manuscript will be deposited in a Zenodo repository upon publication. In the meantime, they are available from the authors upon request.
% =========================================================
\appendix
% =========================================================
\section*{Appendices}
% =========================================================
\section{Noise-averaged potential}
\seclab{Apotential}
% =========================================================

We derive the noise-averaged penalty $\tilde V(u_f) =
\langle V(x(t_f))\rangle$ for the double-well potential \Eqnref{dwell-main},
where $u_f = \langle x(t_f)\rangle$ is the mean final position (we
use $u_f$ throughout this appendix for consistency with \Secref{experiments}).
Under the linearized dynamics~\eqnref{LEQlin} for the harmonic-trap system~\eqnref{meandyn}, the particle position $x(t_f)$ at the final time is
Gaussian with mean $u_f$ and variance $\sigma^2(t_f)$. In the
stationary-variance approximation
$\sigma^2(t_f)\approx\sigma^2_\text{st} = D/|h_{10}| = \kb T/\kappa$,
the distribution of $x(t_f)$ is
\algn{\eqnlab{xfdist}
    x(t_f) \sim \ms{N}(u_f, \kb T/\kappa)\,.
}
We need the first four moments of $x(t_f)$ about the origin. Using the
Gaussian moment formulae $\langle x^2\rangle = u_f^2 + \kb T/\kappa$ and
$\langle x^4\rangle = u_f^4 + 6(\kb T/\kappa)u_f^2 + 3(\kb T/\kappa)^2$,
we obtain
\algn{\eqnlab{moments}
    \langle x^2(t_f)\rangle &= u_f^2 + \frac{\kb T}{\kappa}\,,\nn\\
    \langle x^4(t_f)\rangle &= u_f^4 + 6\frac{\kb T}{\kappa}u_f^2
    + 3\left(\frac{\kb T}{\kappa}\right)^2\,.
}
Substituting into the double-well potential $V_\text{DW}$ in \Eqnref{dwell-main}, we find
\algn{\eqnlab{Vtilderaw}
    \tilde V_\text{DW}(u_f) = \frac{V_0}{4}\left(
    \frac{\langle x^4(t_f)\rangle}{x_\text{m}^4}
    - \frac{2\langle x^2(t_f)\rangle}{x_\text{m}^2} + 1\right)\,.
}
Substituting \Eqnref{moments} into \Eqnref{Vtilderaw} and collecting terms
gives
\algn{\eqnlab{Vtilde}
    \tilde V_\text{DW}(u_f) = V_\text{DW}(u_f) + \frac{V_0\kb T}{4\kappa x_\text{m}^2}
    \left(\frac{6u_f^2}{x_\text{m}^2} - 2 + \frac{3\kb T}{\kappa x_\text{m}^2}\right)\,,
}
where $V(u_f)$ is the bare double-well evaluated at the mean. The correction
terms scale as $\eps = \kb T/(\kappa x_\text{m}^2)$ of \Eqnref{epsdef}. Writing \Eqnref{Vtilde}
in terms of $\eps$ gives \Eqnref{VtildeDW} in the main text.
% =========================================================
\section{Asymmetric obstacles}
\seclab{Aasym}
% =========================================================

Throughout \Secref{transition} we assumed that the noise-averaged
penalty $\tilde V$ is symmetric about the zero-control position. Here we relax
this assumption and determine when a continuous transition survives.

To understand why asymmetry matters, we complete the square in \Eqsref{C0}, \eqnref{Cminus}, and \eqnref{Cplus},
and find that the cost is a parabola centered at some position $\ms{U}^\eta(t_f)$, plus the final cost,
\algn{\eqnlab{Fasym}
	\ms{C}^\eta_{t_f}(u_f, u_0) = \frac{\kappa\Phi^\eta(t_f)}{2}\left[u_f - \ms{U}^\eta(t_f)\right]^2
	+ \tilde V(u_f) + \text{const}\,,
}
with
\sbeqs{\eqnlab{ccentre}
\algn{
	\ms{U}^0(t_f) &= u_0\,,\\
	\ms{U}^-(t_f) &= \frac{u_0}{\cosh(t_f/\tau_\text{c})}\,,\\
	\ms{U}^+(t_f) &= \frac{u_0}{\cos(t_f/\tau_\text{c})}\,,
}
}
for the parabolic, hyperbolic and elliptic classes, respectively.
Only in the parabolic class $\eta=0$ is $\ms{U}^0(t_f)$ independent of $t_f$. There, the running cost
depends on $u_f$ and $u_0$ only through their difference $u_f-u_0$, so one may shift both
endpoints and place the local maximum of $\tilde V$ at the zero-control position without
loss of generality~\cite{Mei26b}. In the other two classes, the cost is anchored to
the zero-control position, while $\tilde V$ is anchored to its local maximum, the obstacle. The
offset between the two cannot be removed in general. Measuring $u_f$ from the local maximum of
$\tilde V$, the offset produces a term linear in $u_f$, which acts in a similar
way as a non-zero $u_0$ in \Secref{landau}, i.e.\ as the analogue of an external
field in Table~\ref{tab:landau}.

\paragraph*{Conditions for a continuous transition.}
Suppressing the equivalence class, we write $F_{t_f,u_0}(u_f) \equiv \ms{C}^\eta_{t_f}(u_f,u_0)$. A continuous transition, characterized by the critical point $(t^\eta_\text{c},u^\text{c}_0,u_f^\text{c})$,
requires a stationary point in $u_f$ at which the curvature and the third derivative
vanish simultaneously,
\algn{\eqnlab{critcond}
	F'_{t_\text{c},u^\text{c}_0}(u_f^\text{c}) = F''_{t_\text{c},u^\text{c}_0}(u_f^\text{c}) = F'''_{t_\text{c},u^\text{c}_0}(u_f^\text{c}) = 0\,.
}
The additional condition $F''''_{t_\text{c},u^\text{c}_0}(u_f^\text{c}) > 0$ ensures that $F_{t_\text{c},u^\text{c}_0}(u_f)$ is binding. Because the control contribution to \Eqnref{Fasym} is quadratic, the third and fourth derivatives involve only the final cost,
$F'''_{t_\text{c},u^\text{c}_0} = \tilde V'''$ and $F''''_{t_\text{c},u^\text{c}_0} = \tilde V''''$, and the conditions
\eqnref{critcond} can be solved one after the other:
\sbeqs{\eqnlab{critsol}
\algn{
	\tilde V'''(u_f^\text{c}) &= 0\,,\\
	\Phi^\eta(t^\eta_\text{c}) &= -\frac{\tilde V''(u_f^\text{c})}{\kappa}\,,\eqnlab{Landau2}\\
	\ms{U}^\eta(t^\eta_\text{c}) &= u_f^\text{c} - \frac{\tilde V'(u_f^\text{c})}{\tilde V''(u_f^\text{c})}\,.
}
}
The first equation locates $u_f^\text{c}$ from the final cost. The
second then fixes the critical time $t^\eta_\text{c}$, and is the analogue of \Eqnref{tcdef} with $u_f^\text{c}$ in
place of $u_f=0$. As before, \Eqnref{Landau2} has a solution only if
$\Phi^\eta$ attains the required value, which for the parabolic and hyperbolic classes
demands conditions analogous to \Eqsref{condition} and \eqnref{condition2}, respectively. The third equation in \Eqsref{critsol} fixes $\ms{U}^\eta(t^\eta_\text{c})$, and hence, through
\Eqnref{ccentre}, the initial position $u_0^\text{c}$ at the critical point. For even $\tilde V$ one has $\tilde V'''(0) = \tilde V'(0) = 0$, so
that $u_f^\text{c}=0$ and $u_0^\text{c} = 0$, and \Eqsref{critsol} reduce to the results of
\Secref{stability}.

\paragraph*{First-order line.}
From the critical $(t^\eta_\text{c},u^\text{c}_0,u_f^\text{c})$ emerges a first-order line $(\tilde t,\tilde u_0,\tilde u_f)$ across which the transition is discontinuous. On this line, $F_{\tilde t,\tilde u_0}(u_f)$ has two degenerate global minima, which implies
\algn{\eqnlab{firstorder}
	F'_{\tilde t,\tilde u_0}(\tilde u_f) = F'''_{\tilde t,\tilde u_0}(\tilde u_f) = 0\,, \quad F''_{\tilde t,\tilde u_0}(\tilde u_f)<0\,, \quad F''''_{\tilde t,\tilde u_0}(\tilde u_f)>0\,,
}
which for a quartic $F_{\tilde t,\tilde u_0}$ implies that the two minima lie symmetrically about $\tilde u_f$. This is one condition fewer than \Eqnref{critcond} and therefore defines a line in the space spanned by $t_f$, $u_0$, and $u_f$. Crossing this line in the $(t_f,u_0)$-space, the optimal final position $u_f^*$ jumps between the two minima and the optimal
cost attains a kink. The line terminates at the critical point $(t^\eta_\text{c},u^\text{c}_0,u_f^\text{c})$, where the two minima
merge and the transition becomes continuous. This is the familiar Landau scenario of a line of discontinuous transitions ending in a critical point, where the transition is continuous. For even $\tilde V$ that line is $\tilde u_0 = 0$ with $\tilde t_f>t_\text{c}^\eta$, along which $u_f^*$ jumps between $\pm|u_f^*|$ as described in \Secref{landau}, and its
endpoint is $(t^\eta_\text{c},u^\text{c}_0,u_f^\text{c}) = (t_\text{c}^\eta,0,0)$. An asymmetric obstacle therefore
does not change the structure of the transition, but it detaches the line from
$u_0 = u_f = 0$ and moves the critical point to non-zero $(u^\text{c}_0,u_f^\text{c})$.

% =========================================================
\section{Experimental methods}
\seclab{Aexpmeth}
% =========================================================

\subsection{Optical tweezers setup}
\seclab{Aexpsetup}

The optical tweezers setup uses a 532~nm laser (Coherent Verdi V2) as
the coherent light source. The beam passes through an acousto-optic
deflector (AOD, AA Opto-Electronic DTSXY-400), which allows simultaneous
amplitude modulation and angular deflection. A telescope of two-inch lenses
directs the deflected beam to the back aperture of the microscope objective
(Olympus MPLAPON-Oil 100$\times$, NA$=1.45$), accommodating deflection
angles of up to $\pm 40$~mrad. The objective serves for both optical
trapping and imaging. The laser power is adjusted so that at the highest
AOD setting used in experiments, it reaches a maximum of 160~mW immediately
before the back aperture.

Microscopy videos are recorded with a digital camera (Basler ace 2
a2A3840-45umPRO). The frame rate is 250~Hz for the optimal control experiment and
400~Hz for both relaxation experiments, with the region of interest adapted
to the specific experimental requirements. The sample temperature is
controlled by resistive heating (Okolab H401-T-Penny) applied to both the
sample stage and the objective, and is maintained at $25\pm 0.05\,
\unit{\celsius}$ throughout all experiments. The laser focus can be shifted
within the imaging plane by up to $\pm 19\,\unit{\micro\meter}$.

To suppress AOD-induced artifacts, a zero-intensity interval is inserted
between consecutive trap positions in the control sequence. Calibration
establishes the mapping between AOD control frequencies and trap positions
by sampling a grid across the field of view and assigning time-averaged
particle positions to the corresponding laser focus locations. A continuous
mapping is obtained via interpolation, and the trap stiffness $\kappa_i$
is characterized across the field of view. The correction procedure
described in \Secref{setup} reduces spatial variations in trap stiffness
from $\pm 20\%$ to $\pm 5\%$.

Sample suspensions are prepared by dispersing spherical silica micro
particles (diameter $2.73\pm 0.12\,\unit{\micro\meter}$, microParticles
GmbH) in a water-glycerol mixture at a volumetric ratio of 1:1. Measurement
cells are assembled by filling glass capillaries (inner diameter
$100\,\unit{\micro\meter}$, width $1\,\unit{\milli\meter}$; CM Scientific)
with the prepared suspensions. The open ends of the capillaries are sealed
using a combination of wax and epoxy resin to prevent evaporation. The
samples are allowed to equilibrate within the measurement setup for a
minimum of 60 minutes prior to data acquisition.

\subsection{Optimal control experiment}
\seclab{Aexpcontrol}

After trapping a particle near the midplane of the sample cell, a
calibration measurement is performed as described in Appendix~\secref{Aexpsetup}.
The optimal control protocols for varying initial positions $u_0$ and
protocol durations $t_f$ are computed from \Eqnref{lam0} and implemented
by moving the AOD-controlled trap center $\lambda(t)$ accordingly.

Each experiment consists of both forward and backward implementations of
the protocol to avoid cumulative drift of the trap across the finite AOD
deflection range. Before, between, and after each forward-backward pair,
the trap is held fixed for $3\,\unit{s}$ to ensure equilibration. Each
forward-backward pair is repeated at least 40 times, yielding 80
trajectories per protocol. The stochastic work is evaluated from
\Eqnref{Wdisc} with time intervals of duration $t_f/(N-1) = 0.004\,
\unit{s}$ set by the $250\,\unit{Hz}$ camera acquisition rate.

\subsection{Free relaxation experiment and reweighting}
\seclab{Aexpfreerelax}

In the free relaxation experiment, the trap is repeatedly switched on and
off for equal periods $t_\text{on} = t_\text{off} = 0.5\,\unit{s}$. During
the off period, the particle diffuses freely with diffusion constant
$D\approx 0.022\,\unit{\micro\meter^2/s}$, obtained from the mean-squared
displacement of the free-diffusion segments, giving $\gamma = \kb T/D\approx 0.18\,\unit{\micro N s/m}$ for the
friction coefficient in the free-relaxation experiment. After
$t_\text{off}$, the trap is switched back on to prevent the particle from
leaving the field of view or settling out of the focal plane. Approximately
$2\times 10^3$ such on-off cycles are recorded.

To compute the rate function $\ms{V}^{0,*}_{t_f}(x_f)$, we use translational
and inversion invariance of free diffusion: $P(x_f, t_f|x_0) = P(|x_f -
x_0|, t_f|0)$, so that
\algn{\eqnlab{Vrewt}
    \ms{V}^{0,*}_{t_f}(x_f) = \min_{y}\left\{
    \ms{A}^0_{t_f}(y) + V(x_f + y)\right\}\,,
}
where $y = x_0 - x_f$ is the displacement accumulated over the snippet and
\algn{\eqnlab{Adef}
    \ms{A}^0_{t_f}(y) = \frac{\gamma y^2}{4 t_f}\approx
    -\kb T\log P_s(y, t_f)
}
is the first term of \Eqnref{V0fixed}, i.e. the Onsager--Machlup action of
free diffusion, with $P_s(y, t_f) = P(y, t_f|0)$ the empirical distribution
of endpoints of trajectories starting from zero. Note that $y$, not $x_0$,
is the minimization variable here. In \Eqnref{Vrewt} we have used translational and inversion invariance
to express $\ms{V}^0_{t_f}(x_f, x_0)$ in terms of the
displacement alone. Now, we shift all measured free-diffusion trajectories so
that they start at $x = 0$, and use the Markov property to divide each
trajectory of duration $t_\text{off}$ into overlapping snippets of the
required duration $t_f$, increasing the effective sample size to $\approx
10^6$ snippets. The resulting empirical distribution $P_s(x_0, t_f)$ is
estimated from these snippets and used to evaluate \Eqnref{Vrewt}.

To compute the order parameter density $P_\text{op}(|x_0|, t_f)=P(|x_0|,0|0,t_f)$ and the trajectory
density of paths conditioned on $x_f = 0$ at time $t_f$, we use the Bayes
relation
\algn{\eqnlab{Bayes}
    P(x_0, 0|0, t_f) = \frac{P(0, t_f|x_0)P(x_0)}{P(0, t_f)}
    \propto P_s(x_0, t_f)\,\ee^{-V_\text{eq}(x_0)/(\kb T)}\,,
}
where we used $P(0, t_f|x_0) = P(x_0, t_f|0) = P_s(x_0, t_f)$ by translational
invariance, and the initial distribution $P(x_0)\propto\ee^{-V(x_0)/(\kb T)}$
is the Boltzmann weight in the double-well. The reweighted distribution
\Eqnref{Bayes} gives the probability that a trajectory ending at $x_f = 0$
at time $t_f$ started at $x_0$. Its concentration around the most probable
initial position $x_0^*$ (the order parameter) is the experimental
signature of the FTDPT.

\subsection{Harmonic relaxation experiment}
\seclab{Aexpharmrelax}

In the harmonic relaxation experiment, the particle relaxes into the
harmonic potential \Eqnref{trap} after a quench. We work in coordinates
shifted so that the trap center is at the origin, consistent with
\Eqnref{trap}. The absence of translational invariance prevents the simple
shifting procedure used in the free relaxation experiment. Instead, we
implement the quench by instantaneously repositioning the trap center
$\lambda$ using a piezo-driven mirror, which achieves repositioning on
timescales much shorter than $\tau_\text{p}^R$.

In the time interval between consecutive mirror jumps ($\approx
0.5\,\unit{s}\approx 7\,\tau_\text{p}^R$), the particle relaxes in the
harmonic potential. We record the particle position throughout these
intervals and divide each trajectory into overlapping snippets of the
required duration $t_f$. Because the post-quench dynamics are
time-homogeneous and Markovian, every such sub-interval is a valid sample of
the relaxation.

However, since $P(x_f, t_f|x_0)\neq P(|x_f - x_0|, t_f|0)$ for $\kappa_\text{q} > 0$,
the snippets cannot be shifted onto the conditioning point as in the free
case. We therefore impose the condition $x_f = 0$ by \emph{selection}: we
retain only those snippets whose endpoint falls inside a narrow window
$|x(t_f)|\leq\delta$ about the trap center, with $\delta = 0.05\,x_\text{m}$. The
results are unchanged within their statistical error for $\delta$ between
$0.02\,x_\text{m}$ and $0.08\,x_\text{m}$.

Selection changes the structure of the reweighting. The retained ensemble is
distributed as $p_\text{exp}(x_0)P(0, t_f|x_0)$. Here, the
distribution $p_\text{exp}$ of snippet start positions carries no physical information but is fixed by the
measurement protocol, including the trap displacement, and by the way the trajectories are cut into snippets. Bayes'
relation \Eqnref{Bayes} therefore requires each snippet to be weighted by
\algn{\eqnlab{harmweight}
    w(x_0) = \frac{\ee^{-V_\text{eq}(x_0)/(\kb T)}}{p_\text{exp}(x_0)}\,,
}
with $p_\text{exp}$ measured from the \emph{unconditioned} snippet ensemble
at time $t_f$. This division has no counterpart in the free-diffusion
case, where the shift makes the empirical distribution of $x_0$ equal to the
propagator factor $P(0, t_f|x_0)$ that \Eqnref{Bayes} requires.

The transition probability $P(x_f, t_f|x_0)$ is estimated from the measured snippets as a two-dimensional histogram
over start and end positions, and the fixed-endpoint rate function
\Eqnref{Vmfixed} follows from $-\kb T\log P(x_f, t_f|x_0)$, normalized at
each $x_f$ by its maximum over $x_0$, which imposes the
property that $\min_{x_0}\ms{V}^-_{t_f}(x_f, x_0) = 0$ for every $x_f$. 

This way, we obtain approximately $1\times 10^7$ overlapping snippets
from $1.2\times 10^5$ relaxation intervals. At $t_f = 1.5\,t_\text{c}^{R,-}$ about $1.6\times 10^6$ of them pass the
selection $|x(t_f)|\leq\delta$.

The critical time $t_\text{c}^{R,-}(\Gamma_\text{DW})$ as a function of
$\Gamma_\text{DW} = V_0/(\kappa_\text{q}x_\text{m}^2)$ is obtained from the same
trajectory dataset without additional measurements, and without repeating
the reweighting for each $\Gamma_\text{DW}$. The transition is located by the
change of sign of the curvature of $\ms{V}^{-,*}_{t_f}$ at $x_f = 0$, which splits into two independent pieces. The first piece is the
curvature of the measured transition rate function
$-\kb T\log P(x_f, t_f|x_0)/V_0$ at $x_f = x_0 = 0$, which in units of
$\kappa_\text{q}$ equals $[\ee^{2t_f/\tau_\text{p}^R}-1]^{-1}$ and is a property of the
dynamics alone, hence independent of $\Gamma_\text{DW}$. The second piece is
the curvature of $V$ at the origin, which contributes $-\Gamma_\text{DW}$ in the same
units. Dividing by $\Gamma_\text{DW}$ returns the dimensionless curvature
$\tilde\Phi^-_R(t_f)$ of \Eqnref{phiR}. Measuring the first piece once, by fitting a parabola
to the transition rate function near the origin at each $t_f$, therefore yields the critical time for \emph{any} $\Gamma_\text{DW}$ from the intersection
\algn{\eqnlab{tcfromcurv}
    \tilde\Phi^-_R\left(t_\text{c}^{R,-}\right) = 1\,.
}
For $\tilde\Phi^-_R(t_f)$ of \Eqnref{phiR}, \Eqnref{tcfromcurv} gives $t_\text{c}^{R,-} = (\tau_\text{p}^R/2)\log(1 + \Gamma_\text{DW}^{-1})$, in agreement with \Eqnref{tcRm} in the main text.
\end{document}